\documentclass[pdflatex,sn-mathphys-num,iicol]{sn-jnl}

\usepackage{amsmath,amssymb,amsfonts}
\usepackage{graphicx}
\usepackage{booktabs}
\usepackage{subcaption}
\usepackage{xspace}
\usepackage{multirow}
\usepackage{xcolor}
\usepackage{float}
\usepackage{enumitem}
\setlist[itemize]{label=$\bullet$}
\usepackage{placeins}

\newcommand{\pt}{p_{\mathrm{T}}\xspace}
\newcommand{\MET}{\ensuremath{E_{\mathrm{T}}^{\mathrm{miss}}}\xspace}
\newcommand{\pquant}{\textsc{PQuantML}\xspace}
\newcommand{\hlsml}{\texttt{hls4ml}\xspace}
\newcommand{\alkaid}{\texttt{Alkaid}\xspace}
\newcommand{\dafml}{\texttt{da4ml}\xspace}
\newcommand{\aiefml}{\texttt{aie4ml}\xspace}
\title[Knowledge Distillation for Real-Time Anomaly Detection]{Knowledge Distillation of a Normalising Flow for Real-Time Anomaly Detection at LHC Level-1 Trigger}

\author[1]{\fnm{Jaiman} \sur{Abson}}

\author[1]{\fnm{Florencia} \sur{Canelli}
\orcid{https://orcid.org/0000-0001-6361-2117}}

\author*[1]{\fnm{Valentina} \sur{Guglielmi}
\orcid{https://orcid.org/0000-0003-3240-7393}}
\email{valentina.guglielmi@physik.uzh.ch}

\author[2]{\fnm{Roope Oskari} \sur{Niemi}\orcid{https://orcid.org/0009-0006-1753-1248}}

\author[2]{\fnm{Maurizio} \sur{Pierini}\orcid{https://orcid.org/0000-0003-1939-4268}}

\author[3]{\fnm{Chang} \sur{Sun}\orcid{https://orcid.org/0000-0003-2774-175X}}

\author[4,5]{\fnm{Francesco} \sur{Vaselli}\orcid{https://orcid.org/0009-0008-8227-0755}}

\affil[1]{\orgdiv{Physik-Institut}, \orgname{Universit\"at Z\"urich (UZH)}, \orgaddress{\city{Z\"urich}, \country{Switzerland}}}
\affil[2]{\orgname{European Organization for Nuclear Research (CERN)}, \orgaddress{\postcode{CH-1211}, \city{Geneva 23}, \country{Switzerland}}}
\affil[3]{\orgname{California Institute of Technology}, \orgaddress{\city{Pasadena}, \state{CA}, \country{United States of America}}}
\affil[4]{\orgname{Scuola Normale Superiore}, \orgaddress{\city{Pisa}, \country{Italy}}}
\affil[5]{\orgname{Istituto Nazionale di Fisica Nucleare (INFN), Sezione di Pisa}, \orgaddress{\city{Pisa}, \country{Italy}}}
\providecommand{\DIFaddend}{} 
\providecommand{\DIFdelFL}[1]{} 
\providecommand{\DIFaddbeginFL}{} 
\providecommand{\DIFaddendFL}{} 
\providecommand{\DIFdel}[1]{} 

\abstract{
We present a new strategy for unsupervised anomaly detection at the hardware-based first stage of event processing (Level-1 trigger) of the Large Hadron Collider (LHC). A normalising flow trained exclusively on Standard Model (SM) events provides a teacher anomaly score based on its exact negative log-likelihood, which is distilled into a compact neural network suitable for field-programmable gate array (FPGA) deployment. The teacher reaches state-of-the-art performance on a benchmark dataset, using a rigorous p-value-based definition of the anomaly score. A simple two-hidden-layer student reproduces the teacher performance across four beyond-the-SM benchmarks to within 0.1 percentage points in area under the receiver operating characteristic curve (AUC), while achieving a compression factor of approximately 325 times. Quantisation-aware training with PQuantML reduces the model to 8-bit precision with AUC changes below 0.06 percentage points. 
Compilation to Verilog firmware with Alkaid yields FPGA designs that require
neither digital signal processors nor block random-access memory and achieve
latencies of 27--52~ns.
This modular pipeline decouples the expressiveness of the anomaly-detection model from real-time hardware constraints, enabling complex architectures to be used for model-agnostic searches for new physics at the full LHC collision rate of 40 MHz.
}


\begin{document}
\maketitle

\section{Introduction}
\label{sec:intro}
Operating at a collision rate of 40\,MHz, the Large Hadron Collider (LHC)
at CERN generates hundreds of terabytes of raw detector data per second in the ATLAS~\cite{ATLAS:2008xda} and CMS~\cite{CMS:2008xjf} experiments alone. Storing this volume in its entirety is not feasible; instead, a tiered trigger system performs online event selection in real time. The first stage, the hardware-based Level-1 Trigger (L1T)~\cite{CMS:2020cmk,CMS:L1TDR}, is implemented on \emph{Field-Programmable Gate Arrays} (FPGAs) and must reduce the incoming rate by roughly three orders of magnitude within a fixed latency of $\mathcal{O}(1\,\mu\mathrm{s})$. Events that survive the L1T are passed to the software-based High-Level Trigger (HLT), which runs on a computer farm and applies more refined reconstruction before writing selected events to disk.

Because the L1T must render its verdict in microseconds, its selection criteria have historically been crafted by hand to match the signatures of known physics processes. This approach proved effective for targeted searches
but is by construction blind to signatures that might not have been anticipated: any new-physics signature failing this first filter is permanently lost to offline analysis. These considerations have motivated a growing body of work to overcome the limitations of the L1T's high rejection rate (${\sim}\,99.75\%$ of all collisions).
Unsupervised machine-learning has emerged as an interesting solution: anomaly detectors~\cite{Kasieczka:2021xcg,Aarrestad:2021oeb} could flag events based solely on their deviation from the learned Standard Model (SM) distribution, without committing to any specific beyond-the-SM (BSM) hypothesis. Early studies applied autoencoders~\cite{Kingma:2013vae} to offline data analysis~\cite{Heimel:2018mkt,Farina:2018fyg}; subsequent proposals extended this paradigm to the trigger, envisioning dedicated data streams for rare or unexpected events~\cite{Cerri:2018anp,Knapp:2020adversarial}.
The first proof of concept for L1T deployment was provided in Ref.~\cite{Govorkova:2021utb}, which demonstrated that a variational autoencoder (VAE) trained on SM data could be synthesised onto an FPGA within the L1T constraints.
The CMS Collaboration subsequently deployed two autoencoder-based anomaly-detection algorithms in 2023: AXOL1TL~\cite{Gandrakota:2024axoltl}, a fully connected VAE operating on high-level features, and CICADA~\cite{Gandrakota:2024axoltl}, a convolutional autoencoder compressed to a compact multilayer perceptron (MLP) via knowledge distillation~\cite{Hinton:2015distill,Pol:2023ots}. In 2025, the ATLAS experiment followed with GELATO~\cite{Cohen:2025gelato}, conceptually similar to AXOL1TL. All algorithms deployed at the LHC to date have used autoencoders, whose anomaly scores rely on the latent-space representation~\cite{Fraser:2021lxm} or on reconstruction error. More recently, Ref.~\cite{Vaselli:2025fad} demonstrated the first application of Continuous Normalising Flows (CNFs) for L1T anomaly detection. That work introduced a novel hardware-friendly anomaly score defined as the squared norm of the model's vector field output, circumventing the need for computationally expensive \emph{ordinary differential equation} (ODE) integration on an FPGA, and showed that normalising flows (NFs) can achieve performance comparable to autoencoders while meeting the stringent latency and resource constraints of the L1T.

In this work, we pursue a complementary approach that leverages the strengths of both expressive generative models and compact hardware-deployable networks. Rather than designing a new model architecture or anomaly score for direct FPGA deployment, we use \emph{knowledge distillation}~\cite{Pol:2023ots} to compress the anomaly scores of a powerful but computationally prohibitive NF \emph{teacher} into a compact \emph{student} network suitable for FPGA implementation, similarly to what was done for CICADA. A key advantage of this choice is that NFs provide an \emph{exact} and tractable log-likelihood, as opposed to the  lower bound (ELBO) approximation used by variational autoencoders; one can use the p-value of a given event as the distillation target, reflecting how well each event is genuinely described by the learned SM density.
The use of knowledge distillation offers a decoupling between model expressiveness and hardware constraints~\cite{Pol:2023ots}. This  gives access to NFs and similar powerful anomaly-detection strategies  without facing the limitations imposed by the strict latency and resource budget of the L1T.

The main contributions of this work are:
\begin{itemize}[leftmargin=2em]
  \item We train a NF teacher on low-level SM features, using the negative log-likelihood as an anomaly score;
  \item We evaluate the performance on four new physics benchmarks at floating-point precision;
  \item We distil the teacher's output into a compact MLP student  with negligible loss in anomaly-detection performance across all four BSM benchmarks;
  \item We compress the student model through  quantisation-aware training (QAT) using the \pquant framework~\cite{Niemi_2026}, and perform a systematic bit-width scan demonstrating that 8-bit precision preserves the full-precision performance while 3-bit precision remains functional with a moderate loss;
  \item We convert the quantised models into register-transfer-level (RTL) designs for an FPGA using the \alkaid library~\cite{Sun:2025orx}, establishing a complete pipeline from teacher training to hardware-deployable firmware.
\end{itemize}

The remainder of this paper is organised as follows. Section~\ref{sec:data} describes the data samples used for training and evaluation. Section~\ref{sec:teacher} presents the teacher model and its performance. Section~\ref{sec:student} describes the knowledge distillation procedure and the student model. Section~\ref{sec:quantisation} details the quantisation-aware training and the results of the bit-width scan. Section~\ref{sec:fpga} covers the direct-RTL FPGA implementation with \alkaid. Section~\ref{sec:discussion} provides a comparative discussion with existing alternative methods and Section~\ref{sec:conclusions} summarises our conclusions.

\section{Data samples}
\label{sec:data}
This study uses the publicly available LHC benchmark datasets introduced in Ref.~\cite{Govorkova:2021utb} and released on Zenodo~\cite{zenodo:lq,zenodo:a4l,zenodo:h0tt,zenodo:hctaunu}; their composition and generation are documented in Ref.~\cite{Govorkova:2021dataset}. Using the same datasets as Ref.~\cite{Vaselli:2025fad} allows a direct comparison against the CNF-based approach.

The SM background sample~\cite{Aarrestad:2021v2} consists of simulated proton--proton collision events that have been pre-selected by requiring at least one electron with $\pt > 23$\,GeV and $|\eta| < 3$, or at least one muon with $\pt > 23$\,GeV and $|\eta| < 2.1$. This kinematic selection mimics the acceptance of a typical all-purpose L1T algorithm.
We use 3.5 million events, of which 2.8 million are used for training and 700{,}000 for validation. An independent sample of 2 million events is used for testing.
To assess out-of-distribution sensitivity, we adopt the four BSM signal benchmarks originally introduced in Ref.~\cite{Cerri:2018anp}:
\begin{itemize}[leftmargin=2em]
  \item A leptoquark ($LQ$, $m = 80$\,GeV) decaying to a $b$ quark and a $\tau$ lepton: $LQ \to b\tau$~\cite{zenodo:lq};
  \item A neutral scalar ($A$, $m = 50$\,GeV) decaying via two off-shell Z bosons to four leptons: $A \to 4\ell$~\cite{zenodo:a4l};
  \item A neutral scalar ($h^0$, $m = 60$\,GeV) decaying to a $\tau$ pair: $h^0 \to \tau\tau$~\cite{zenodo:h0tt};
  \item A charged scalar ($h^\pm$, $m = 60$\,GeV) decaying to a $\tau$ lepton and a neutrino: $h^\pm \to \tau\nu$~\cite{zenodo:hctaunu}.
\end{itemize}
The available event statistics for the BSM benchmarks are 340,544 ($LQ$), 55,969 ($A \to 4\ell$), 691,283 ($h^0 \to \tau\tau$), and 760,272 ($h^\pm \to \tau\nu$).
None of these signal events enter the training or validation sets; they serve exclusively as post-training performance benchmarks.

Each event is represented by three kinematic quantities (transverse momentum $\pt$, pseudorapidity $\eta$, and azimuthal angle $\phi$) for each of 18 reconstructed objects (four muons, four electrons, ten jets). Additionally, the missing transverse energy (\MET) is described by its magnitude $|\MET|$ and azimuthal angle $\phi_{\mathrm{MET}}$, with the pseudorapidity $\eta$ set to zero by construction. Events containing fewer than the maximum multiplicity of a given object type are zero-padded to maintain a fixed input shape, as done in Refs.~\cite{Govorkova:2021utb,Vaselli:2025fad} and consistent with how the L1T operates. Flattening this object list yields a 57-dimensional input vector for every event. The inputs are standardised via standard scaling during both training and inference.

\section{The teacher flow model}
\label{sec:teacher}

A NF~\cite{Papamakarios:2019fms} is a generative model that constructs a tractable density estimate of the training data by learning a bijection $f: \mathbb{R}^D \to \mathbb{R}^D$ between the data space $\mathbf{x}$ and a latent space $\mathbf{z}$ equipped with a simple prior $p(\mathbf{z})$ (typically a standard Gaussian):
\begin{equation}
  \mathbf{x} = f(\mathbf{z}) \quad \text{and} \quad \mathbf{z} = f^{-1}(\mathbf{x}) .
\end{equation}
Because $f$ is required to be invertible and differentiable, the data density can be evaluated exactly via the change-of-variables formula:
\begin{equation}
  \log p(\mathbf{x}) = \log p\!\left(f^{-1}(\mathbf{x})\right) + \log \left| \det \frac{\partial f^{-1}}{\partial \mathbf{x}} \right| .
  \label{eq:change_of_var}
\end{equation}
This exact and tractable likelihood distinguishes NFs from other deep generative models with approximate likelihoods, such as VAEs. By matching the input distribution to a Gaussian distribution in a latent space, one can associate each point of the input space 
to the p-value of the corresponding latent space. Any monotonic function of this p-value could be used to define an anomaly score.
As a result, NFs provide a natural framework for density-based anomaly detection, where events poorly described by the learned SM distribution are expected to exhibit low likelihood (or equivalently high negative log-likelihood), without requiring an explicit BSM hypothesis. This interpretation, however, is generally less robust for models with approximate likelihoods.

Real-valued Non-Volume Preserving (RealNVP)~\cite{Dinh:2016realnvp} flows are a special kind of flow which constructs the invertible mapping $f$ as a composition of $L$ affine coupling layers. Each coupling layer partitions the input $\mathbf{x}$ into two subsets using a binary mask $\mathbf{m}$:
\begin{align}
  \mathbf{x}_{\mathrm{masked}} & = \mathbf{m} \odot \mathbf{x} ,                                                                                                                                             \\
  \mathbf{x}' & = (1 - \mathbf{m}) \odot
  \left[ \mathbf{x} \odot \exp\bigl(s(\mathbf{x}_{\mathrm{masked}})\bigr) \right. \notag\\
  &\quad \left. +\, t(\mathbf{x}_{\mathrm{masked}}) \right]
  + \mathbf{x}_{\mathrm{masked}} ,
\end{align}
where $s(\cdot)$ and $t(\cdot)$ are the scale and translation functions parameterised by neural networks. The log-determinant of the Jacobian takes the computationally efficient form $\sum_j s_j(\mathbf{x}_{\mathrm{masked}})$.
Our teacher model uses $L = 8$ coupling layers with alternating binary checkerboard masks (alternating between even- and odd-indexed dimensions across layers), ensuring that all 57 input features are transformed across the full stack.
Each coupling layer contains two independent sub-networks for $s(\cdot)$ and $t(\cdot)$, each consisting of three hidden layers of 256 units with rectified linear unit (ReLU) activations. 
The scale network uses a $\tanh$ output activation scaled by a factor of 0.5 to bound the magnitude of the scale transformation and stabilise training. The translation network uses a linear output activation. The base distribution is a 57-dimensional standard multivariate normal, $\mathcal{N}(\mathbf{0}, \mathbf{I}_{57})$.
The total number of trainable parameters is approximately 2.6M.

\subsection{Training and anomaly score}
The teacher is trained by minimising the mean value of the negative log-likelihood (NLL) $\ell(\mathbf{x}) = -\log p(\mathbf{x})$ over the SM training set:
\begin{align}
  \mathcal{L}_{\mathrm{teacher}} = \frac{1}{N} \sum_{i=1}^{N} \ell(\mathbf{x}_i)
  &= -\frac{1}{N} \sum_{i=1}^{N} \left[ \log p(\mathbf{z}_i) \right. \notag\\
  &\quad \left. +\sum_{l=1}^{L} \log \left| \det
  \frac{\partial f_l^{-1}}{\partial \mathbf{x}_i} \right| \right] .
\end{align}
where the sum runs over all $N$ SM training events. The model is optimised with the Adam optimiser~\cite{Kingma:2014adam} with a learning rate of $10^{-4}$ and a batch size of 4096, for up to 200 epochs with early stopping (patience of 5 epochs on the validation loss, restoring the best weights).
After training, the per-event NLL serves as the anomaly score: events with higher NLL are less compatible with the learned SM distribution and are flagged as more anomalous.

\begin{figure}[ht!]
  \centering
   \DIFaddbeginFL \includegraphics[width=\columnwidth]{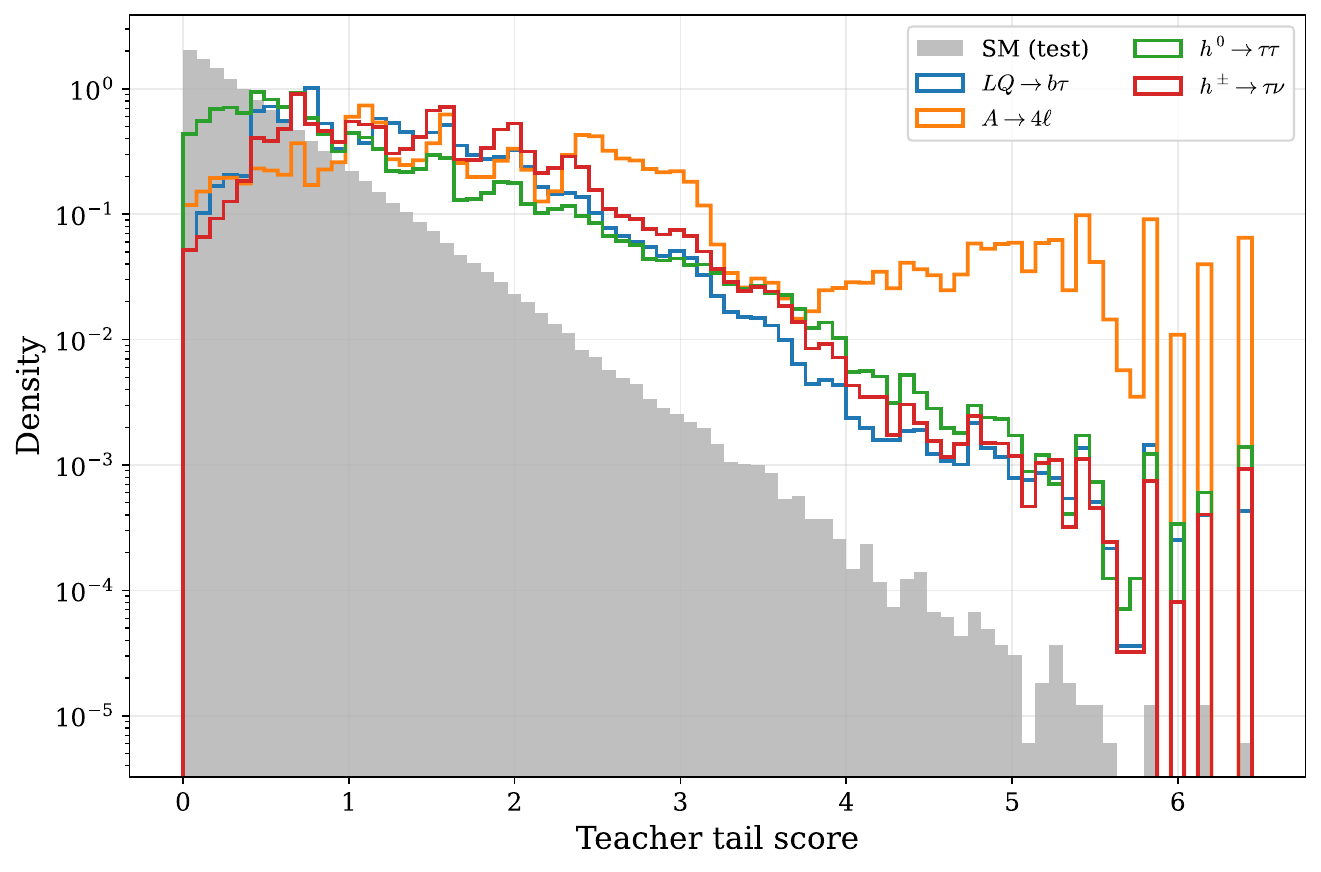}
  \DIFaddendFL \caption{Teacher score distributions after applying the tail-aware ECDF
transformation of Eq.~\eqref{eq:ECDF} to the teacher NLL, for SM test samples and the four anomaly classes}
  \label{fig:scores1}
\end{figure}

To produce a tail-aware regression target for the student, we apply an
empirical quantile transform to the teacher NLL. Each event is assigned its
empirical cumulative distribution function (ECDF) value, computed from the
SM training sample, and the upper-tail probability is mapped to a logarithmic
scale. We define the transformed score $s(\ell)$ as
\begin{multline}
\label{eq:ECDF}
s(\ell)=-\log_{10}\left[1-\frac{1}{N+1}\right.\\
\left.\times\left(\min\!\left(
\left|\left\{i:\ell_i^{\mathrm{train}}<\ell\right\}\right|,N-1
\right)+1\right)\right],
\end{multline}
where $N$ is the number of SM training events. The clipping of the empirical
rank at $N-1$ prevents the argument of the logarithm from vanishing for events
with an NLL larger than all values in the training sample.
This monotone non-decreasing transformation preserves the relative ordering
of the events while giving greater resolution to the high-NLL tail. Under
this mapping, increasingly anomalous events receive progressively larger
scores, making the target particularly sensitive to the low-background
region relevant for fixed-FPR working points. 
Figure~\ref{fig:scores1} shows the resulting teacher tail-score distributions for the SM test sample and the four anomaly benchmarks.

\subsection{Teacher performance}
\label{sec:teacher_performance}

\begin{figure}[t]
  \centering
   \DIFaddbeginFL \includegraphics[width=\columnwidth]{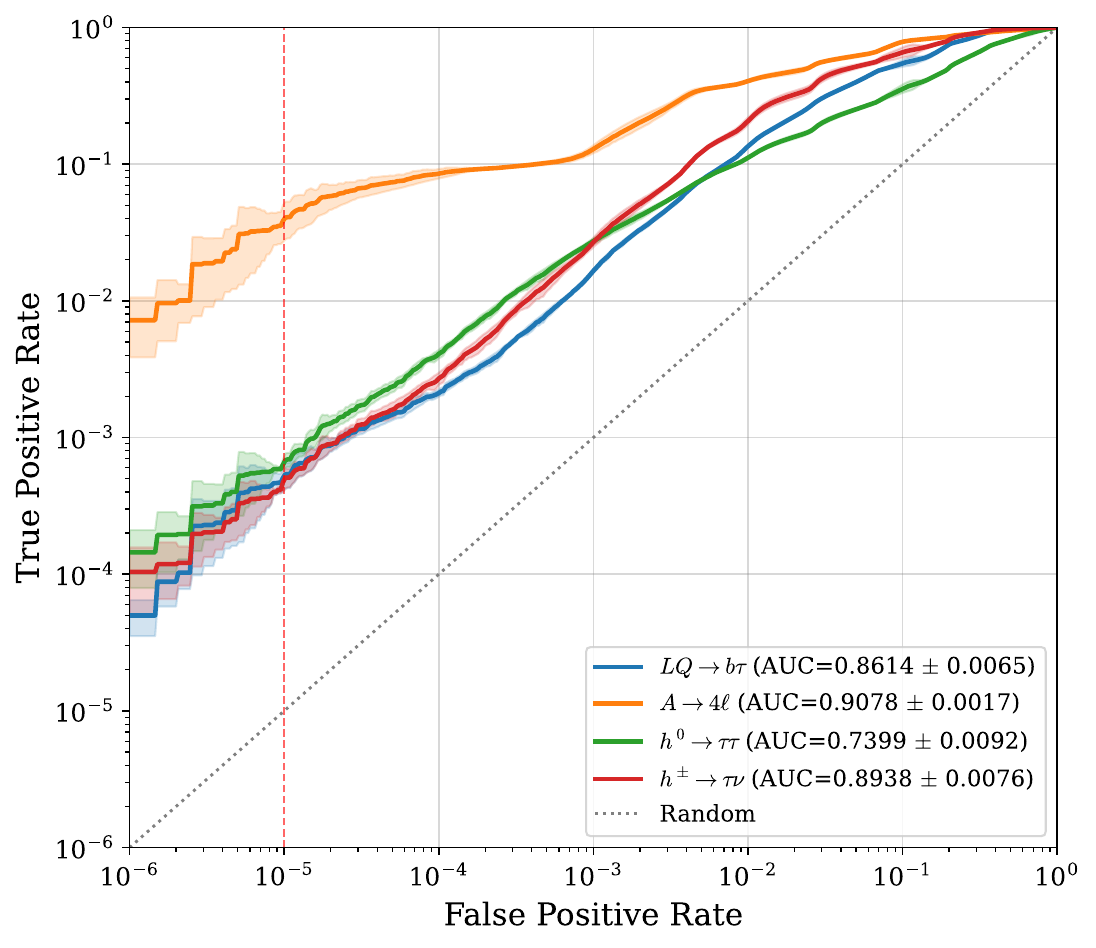}
  \DIFaddendFL \caption{ROC curves for the RealNVP teacher model using the NLL anomaly score, evaluated on the four BSM benchmark signals against the SM test set. Curves correspond to the nominal result and shaded bands show the RMS deviations of four additional independent trainings  performed using different random seeds}
  \label{fig:teacher_roc}
\end{figure}

The performance of the teacher model is evaluated using receiver operating characteristic (ROC)
curves, which display the true positive rate (TPR) as a function of the false positive rate (FPR) as the threshold on the anomaly
score is varied. The ROC curves and corresponding area under the curve (AUC) values are shown in
Figure~\ref{fig:teacher_roc}. 
To assess the robustness of the teacher performance against training stochasticity, we repeat the complete training procedure using five independent random seeds. Each seed determines the random partition of the 3.5 million SM training events into training and validation subsets, as well as the model weight initialisation and mini-batch ordering. The same statistically independent test sample is used to evaluate all trained models, ensuring a consistent comparison across seeds.

In order to estimate the uncertainty associated with the choice of the training starting conditions, we vary the seed with respect to the nominal value. The quoted uncertainties are the root-mean-square (RMS) deviations of the results obtained with the four alternative seeds from the nominal result.
The teacher achieves strong discrimination
across all four BSM benchmarks: the highest AUC is achieved for the
$A \to 4\ell$ signal (90.78 \%), while the lowest is found for
$h^0 \to \tau\tau$ (73.99\%), which is kinematically more similar to the
SM background and therefore harder to distinguish.
The performance is further quantified in Table~\ref{tab:TPR}, which reports the AUC together with the TPR at fixed FPR working points  from $10^{-2}$  to $10^{-5}$. With two million events in the SM test sample,  these working points correspond nominally to approximately 20,000, 2,000, 200, and 20 background events, respectively. \footnote{For the teacher model, all AUC and TPR values reported throughout this paper are computed directly
from the raw NLL score.}

\begin{table*}[t]
  \centering
  \caption{Performance of the teacher model across the four BSM benchmarks using the NLL anomaly score. Values correspond to the nominal training; uncertainties represent the RMS deviations of the results from four additional independent trainings, performed with different random seeds, relative to the nominal result}
  \label{tab:TPR}
   \DIFaddbeginFL \resizebox{\textwidth}{!}{%
  \begin{tabular}{lrrrrr}
    \toprule
                        & AUC (\%) & TPR @ $10^{-2}$ (\%) & TPR @ $10^{-3}$ (\%) & TPR @ $10^{-4}$ (\%) & TPR @ $10^{-5}$ (\%) \\
    \midrule
    LQ$\to b\tau$       & $86.14\pm0.65$ & $13.52\pm0.15$ & $1.66\pm0.05$ & $0.21\pm0.01$ & $0.05\pm0.01$ \\
    A$\to 4\ell$        & $90.78\pm0.17$ & $40.52\pm0.91$ & $12.91\pm0.85$ & $8.54\pm0.69$ & $4.07\pm1.25$ \\
    $h^0\to\tau\tau$    & $73.99\pm0.92$ & $11.13\pm0.25$ & $2.75\pm0.13$ & $0.41\pm0.03$ & $0.07\pm0.01$ \\
    $h^{\pm}\to\tau\nu$ & $89.38\pm0.76$ & $20.66\pm1.27$ & $2.70\pm0.21$ & $0.27\pm0.03$ & $0.05\pm0.01$ \\
    \bottomrule
  \end{tabular}%
  }
\DIFaddendFL \end{table*}


\section{Knowledge distillation and student model}
\label{sec:student}

The RealNVP teacher, with approximately 2.6M parameters distributed across 8 coupling layers each containing two 3-layer MLPs, is far too large and computationally intensive for direct deployment on an FPGA within the sub-microsecond latency budget of a typical LHC L1T. Knowledge distillation~\cite{Hinton:2015distill} provides a principled framework for transferring the teacher's learned mapping into a much smaller network.
The key insight is that reproducing a known anomaly-detection strategy in a supervised training is substantially easier than discovering such a strategy with an unsupervised one. As a result, the student network can learn to approximate the teacher’s behaviour with significantly lower complexity, while retaining much of its discriminative power~\cite{Pol:2023ots}.

In order to meet the requirements of the current L1T infrastructure, we choose a student network based on a multilayer perceptron architecture. The model consists of two hidden layers with 64 units each and ReLU activations, followed by a single output neuron with linear activation.
With 7,937 trainable parameters, the student achieves an architectural compression factor of approximately $325\times$ relative to the teacher. Further compression can be achieved through quantisation during FPGA deployment, as described in Section~\ref{sec:quantisation}.

\subsection{Training}

The student is trained to regress the teacher’s ECDF-transformed anomaly scores (Eq.~\ref{eq:ECDF}), by minimising the Mean Absolute Error (MAE) loss function.
The choice of MAE over Mean Squared Error (MSE) is motivated by its robustness to the long tail of the NLL distribution.
Training uses the Adam optimiser~\cite{Kingma:2014adam} with an initial learning rate of $10^{-3}$ and weight decay of $10^{-4}$, with a batch size of 1024. A \texttt{ReduceLROnPlateau} schedule (patience 5, factor 0.5, minimum learning rate $10^{-6}$) and early stopping (patience 10, restoring best weights) are employed. Training proceeds for up to 200 epochs. The student is trained exclusively on SM events using the teacher's scores as regression targets.
Similarly to the teacher, the student training is repeated for four additional random seeds to assess the robustness of the distillation performance against variations in the data split, weight initialisation, and mini-batch orde8ring.
Within each training, teacher and student models use the same train/validation/test data.

As an additional robustness check, we explicitly tested whether reusing the teacher-training events for student training biases the distilled model. Keeping the nominal teacher fixed, we trained two otherwise identical students with equal training statistics: one on SM events used to train the teacher and the other on a statistically disjoint SM sample. Both students were evaluated on the same blind SM test set and the same four anomaly benchmarks. Their performances are consistent within the observed seed-to-seed fluctuations, with a maximum absolute AUC difference of $3.4\times10^{-4}$ across the four benchmarks and no systematic advantage for the same-event configuration. We therefore find no evidence that reusing the teacher-training events drives the reported student performance and adopt this configuration for the nominal results.

\subsection{Distillation performance}
\label{sec:distillation_performance}

\begin{figure*}[t]
  \centering
  \begin{subfigure}[t]{0.32\textwidth}
     \DIFaddbeginFL \includegraphics[width=\textwidth]{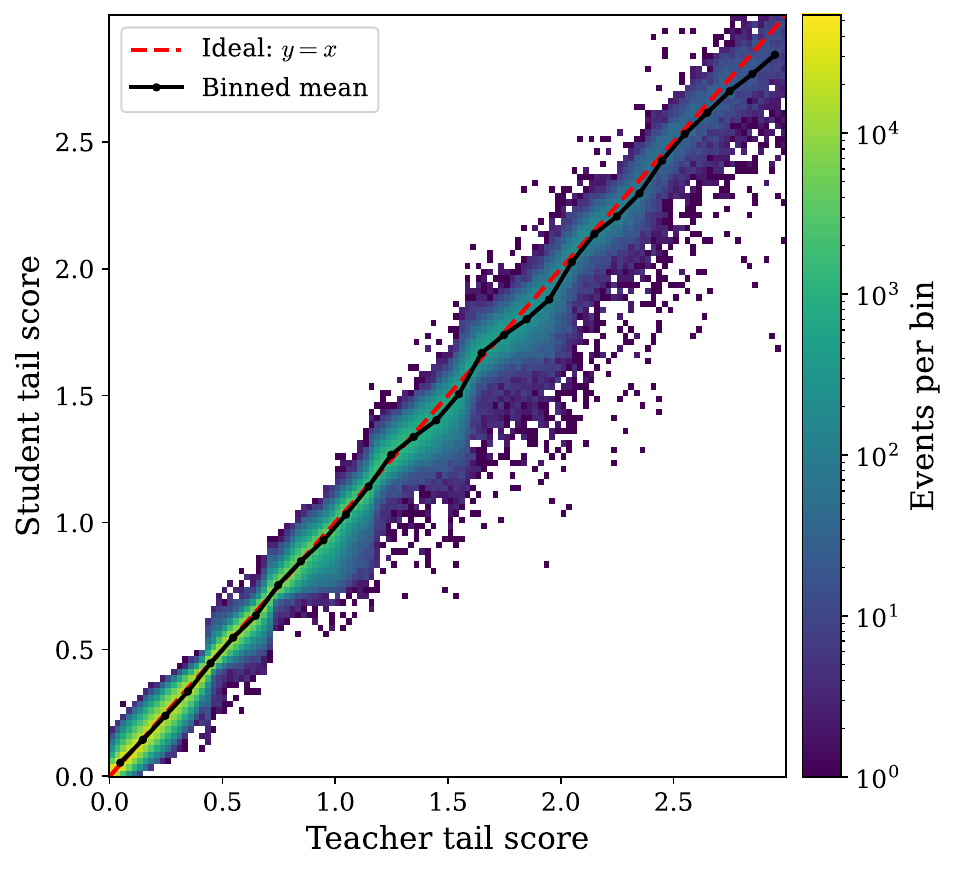}
    \DIFaddendFL \caption{SM test set}
  \end{subfigure}
  \hfill
  \begin{subfigure}[t]{0.32\textwidth}
     \DIFaddbeginFL \includegraphics[width=\textwidth]{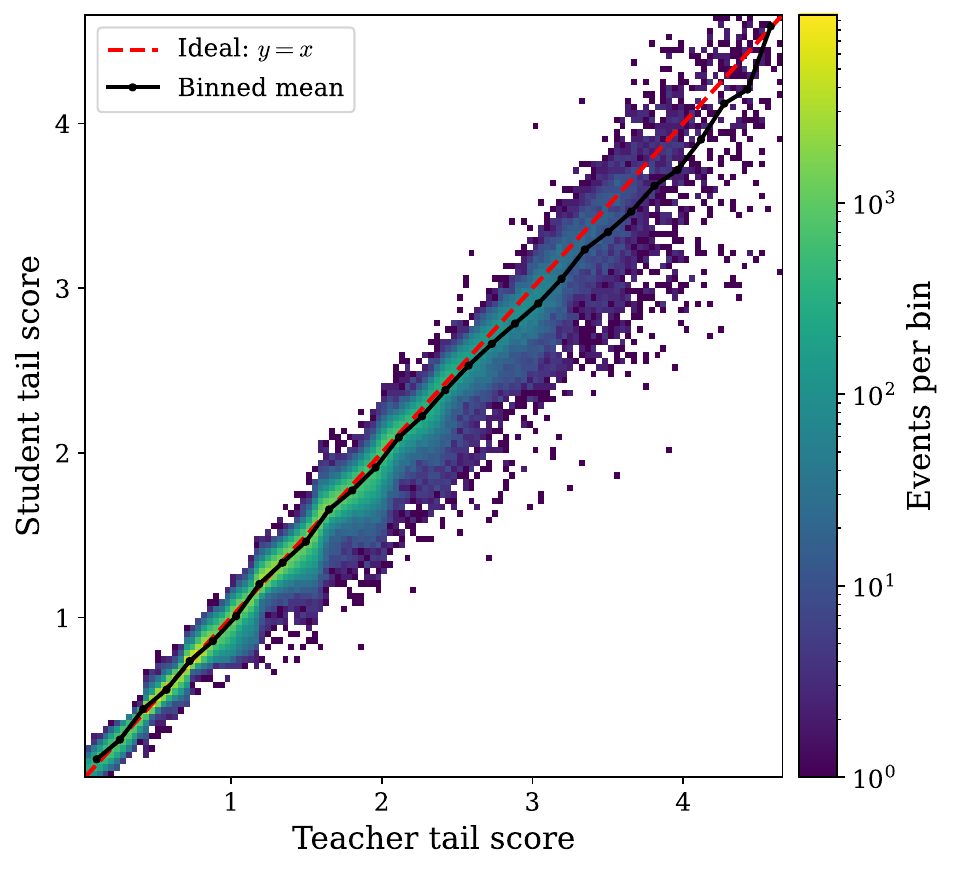}
    \DIFaddendFL \caption{$LQ \to b\tau$}
  \end{subfigure}
  \hfill
  \begin{subfigure}[t]{0.32\textwidth}
     \DIFaddbeginFL \includegraphics[width=\textwidth]{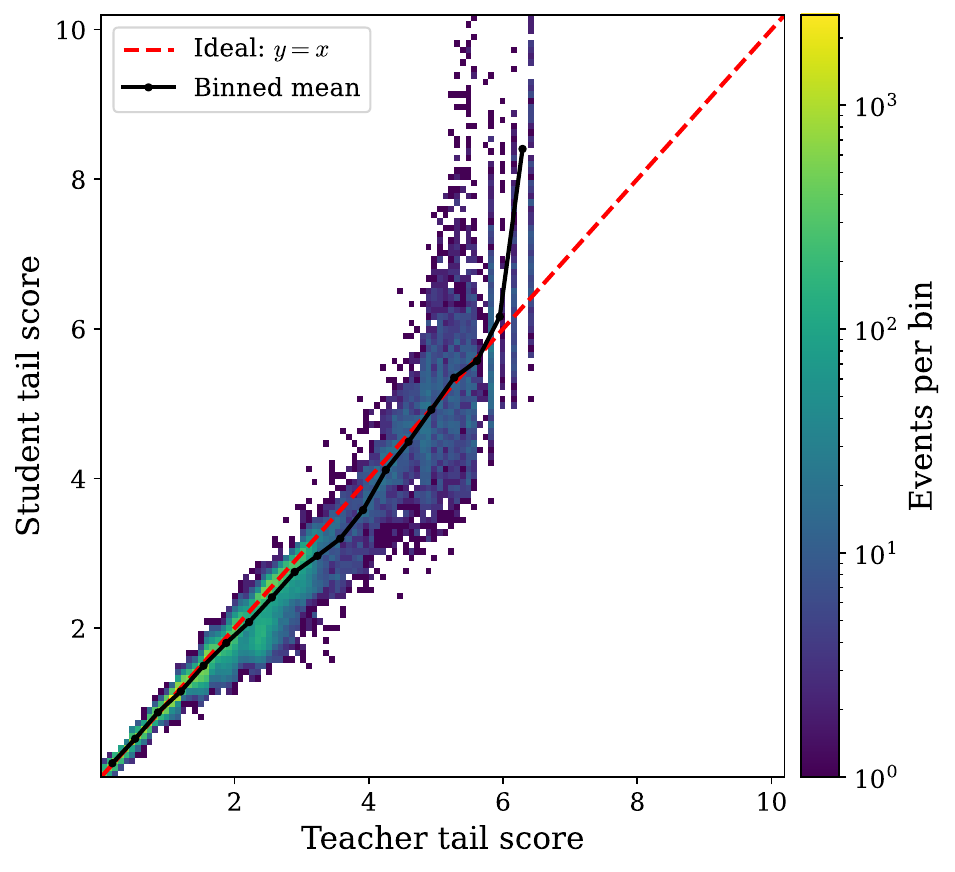}
    \DIFaddendFL \caption{$A \to 4\ell$}
  \end{subfigure}

  \vspace{0.4cm}

  \begin{subfigure}[t]{0.32\textwidth}
     \DIFaddbeginFL \includegraphics[width=\textwidth]{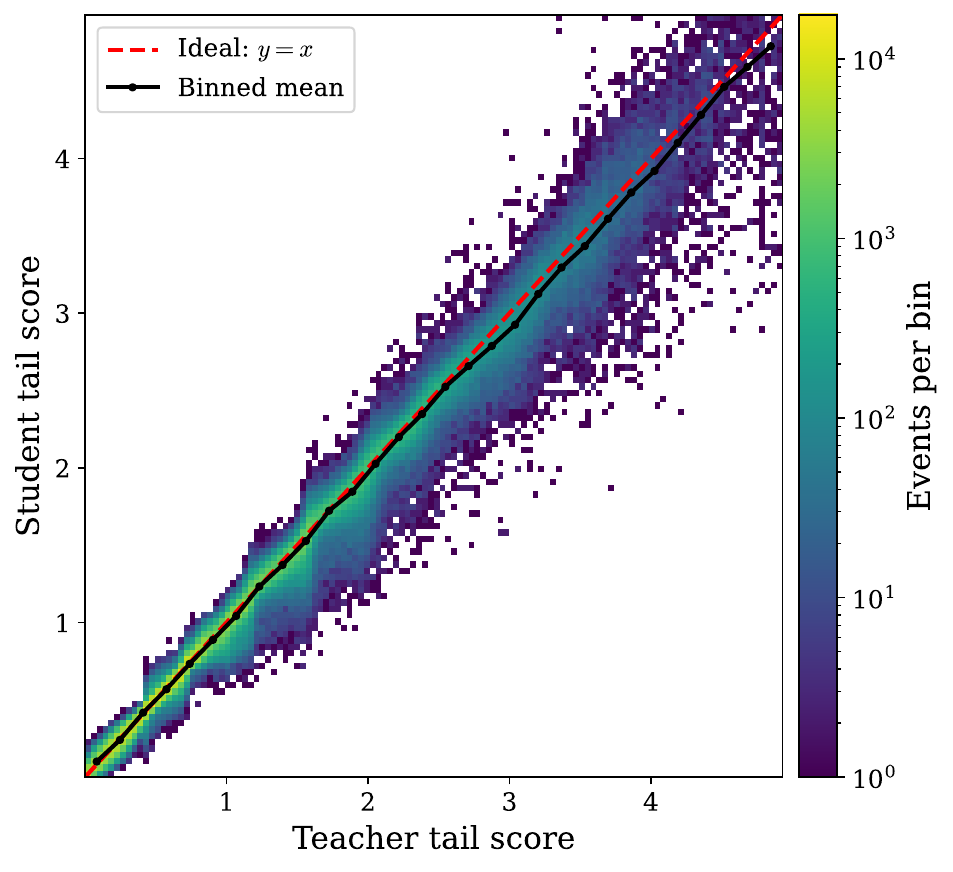}
    \DIFaddendFL \caption{$h^0 \to \tau\tau$}
  \end{subfigure}
  \hfill
  \begin{subfigure}[t]{0.32\textwidth}
     \DIFaddbeginFL \includegraphics[width=\textwidth]{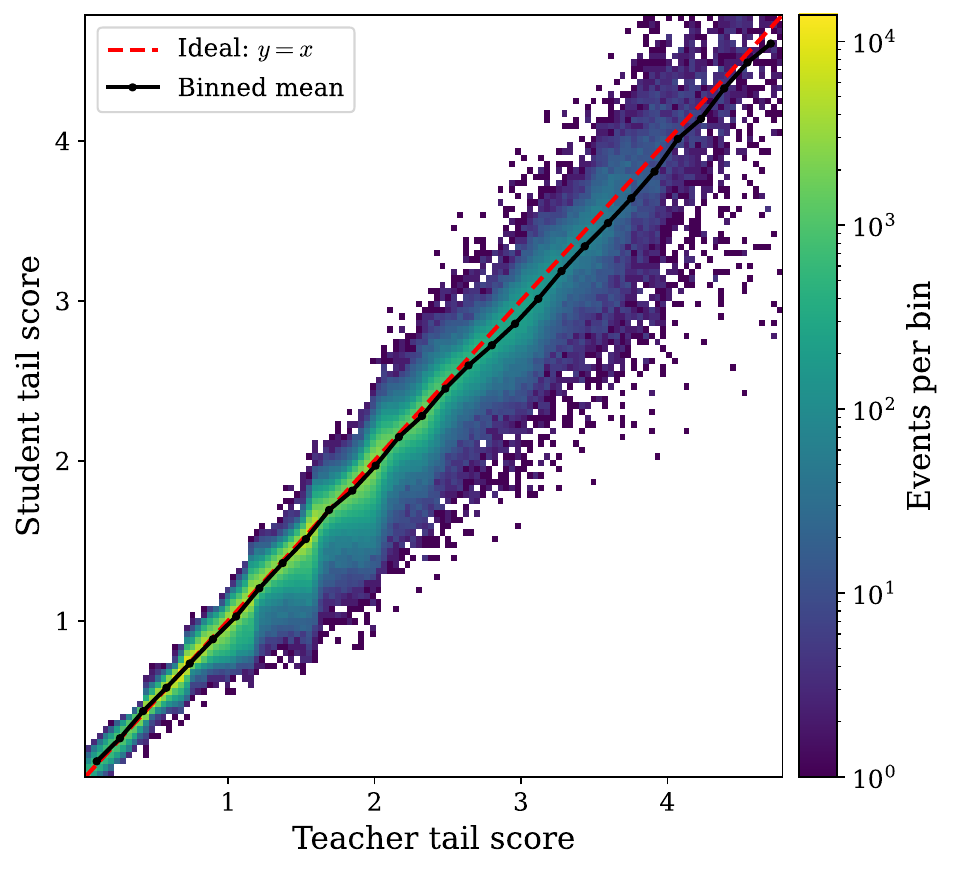}
    \DIFaddendFL \caption{$h^\pm \to \tau\nu$}
  \end{subfigure}
  \hfill
  \begin{subfigure}[t]{0.32\textwidth}
     \DIFaddbeginFL \phantom{\includegraphics[width=\textwidth]{Fig3a.pdf}}
  \DIFaddendFL \end{subfigure}

  \caption{Calibration of student scores with respect to the teacher tail-aware ECDF scores on (a)~the SM test set, (b)~$LQ \to b\tau$, (c)~$A \to 4\ell$, (d)~$h^0 \to \tau\tau$, and (e)~$h^\pm \to \tau\nu$. The solid black line is the binned mean; the dashed red line is $y = x$ (perfect calibration). The binned mean closely follows the diagonal across all datasets, confirming faithful score reproduction. Scatter increases at high scores where events are sparse}
  \label{fig:calibration}
\end{figure*}

To verify that the student faithfully reproduces the teacher's anomaly scores at the event level, we examine the calibration of the student's output against the teacher's ECDF-transformed NLL. Figure~\ref{fig:calibration} shows this comparison on the SM test set and all four BSM benchmarks.
Across all five datasets, the binned mean closely follows the $y=x$ diagonal, confirming that the student faithfully reproduces the teacher scores.
The event-level scatter increases in the high-score tail, where events are
sparse and the regression task is inherently more challenging.

\begin{figure}[h!]
  \centering
   \DIFaddbeginFL \includegraphics[width=\columnwidth]{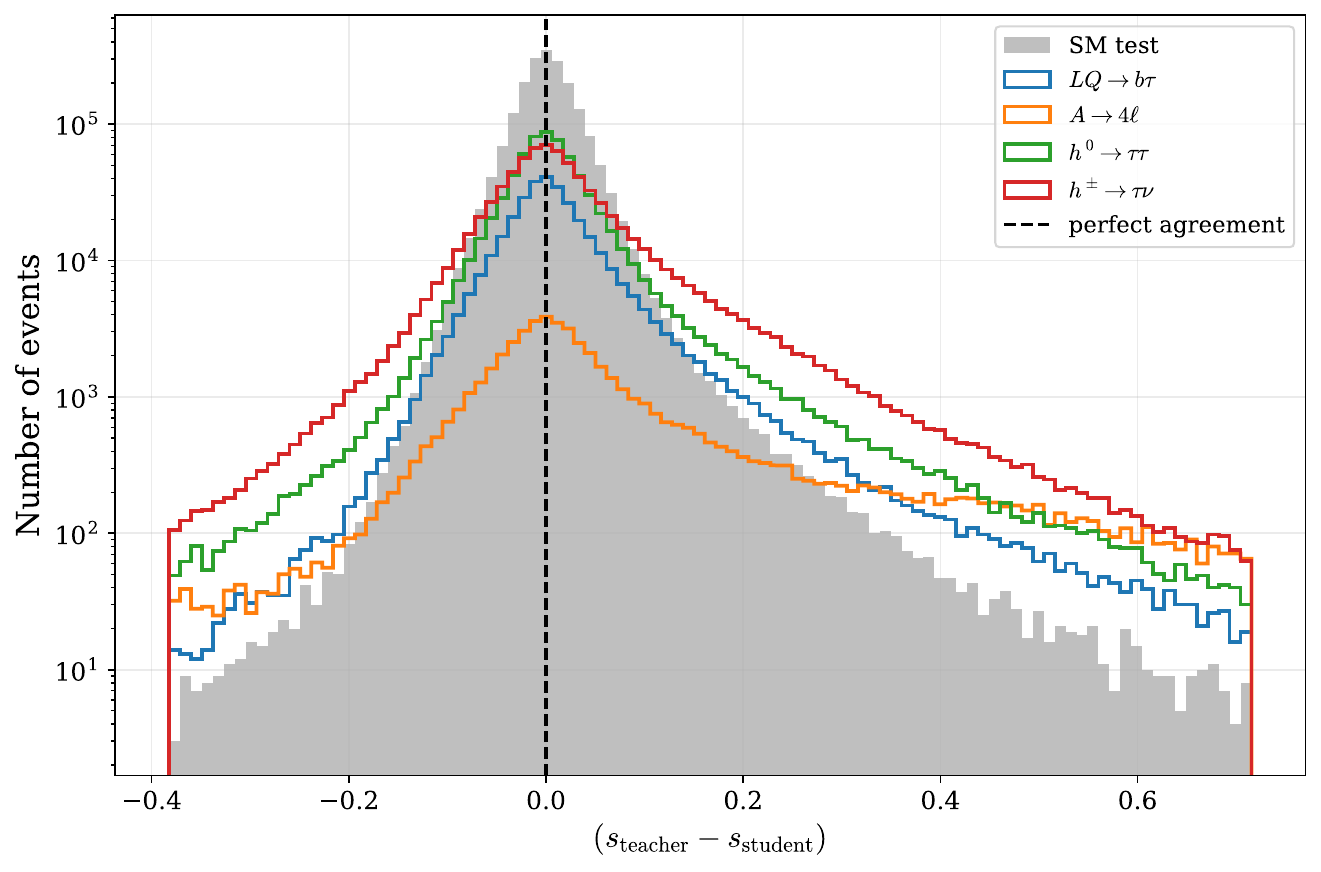}
  \DIFaddendFL \caption{Distributions of the residual $(s_{\mathrm{teacher}} - s_{\mathrm{student}})$ for SM test events and the four anomaly classes. The dashed vertical line at zero indicates perfect agreement between teacher and student scores}
  \label{fig:scores}
\end{figure}

To further quantify the event-by-event agreement between teacher and student predictions, Figure~\ref{fig:scores} shows the distributions of the residual $(s_{\mathrm{teacher}} - s_{\mathrm{student}})$ for the SM test set and all anomaly benchmarks. The residual distributions are strongly peaked around zero for all datasets, demonstrating that the student reproduces the teacher scores with high accuracy on an event-by-event basis. The broader positive tails observed for anomalous events are consistent with the increased scatter at high scores seen in Figure~\ref{fig:calibration}, where the event density becomes sparse.
Notably, the calibration on the BSM benchmarks (which were never seen during training) is comparably tight to that on the SM test set, indicating that the student generalises well to out-of-distribution events.

\begin{figure}[h!]
  \centering
   \DIFaddbeginFL \includegraphics[width=\columnwidth]{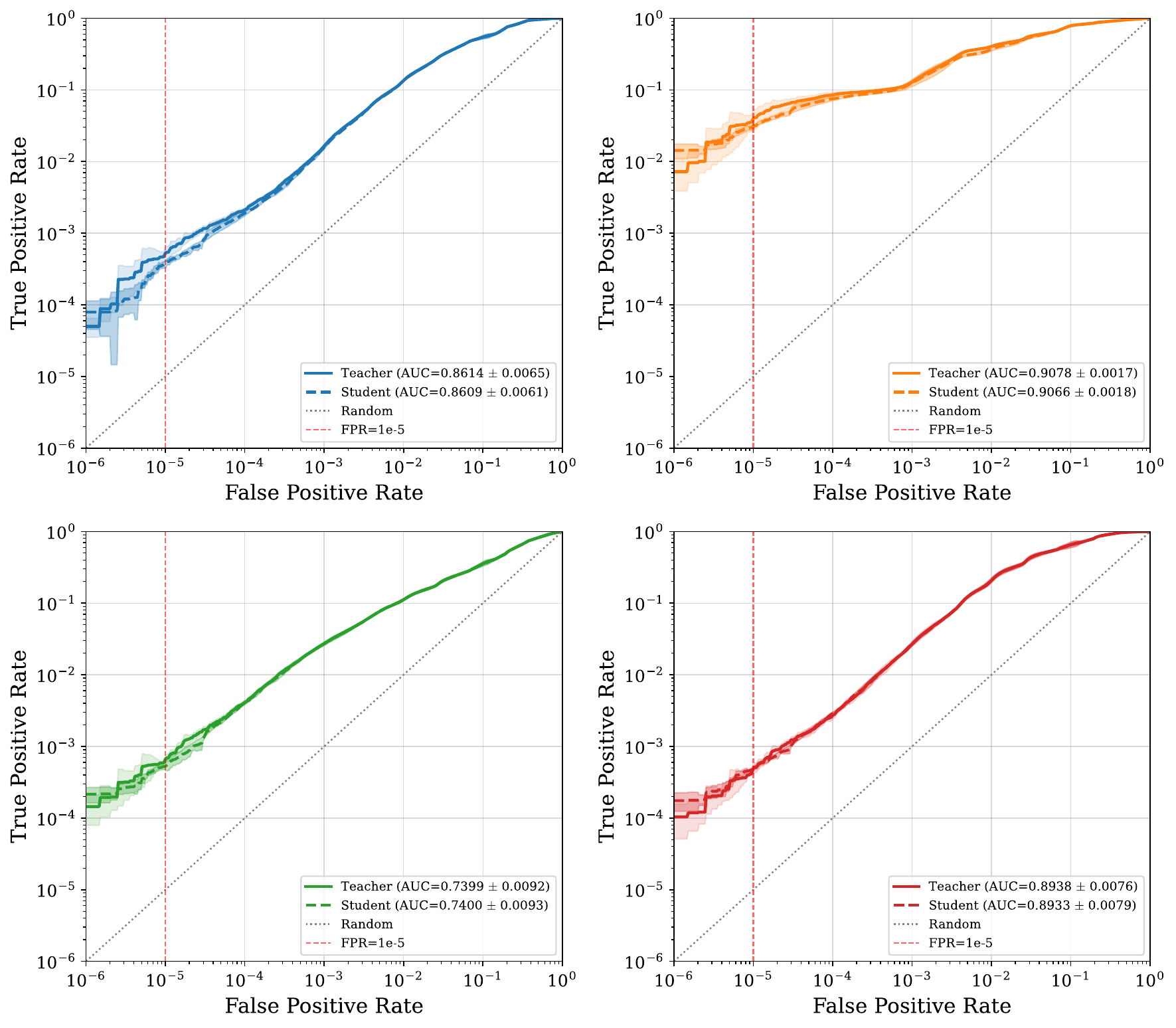}
  \DIFaddendFL \caption{ROC curves comparing the RealNVP teacher (solid) and distilled MLP student (dashed) for the nominal result. The shaded bands represent the RMS deviations of four additional independent trainings, performed using different random seeds, from the nominal curves; the quoted AUC uncertainties are defined in the same way}
  \label{fig:teacher_vs_student_roc}
\end{figure}

Despite the increased scatter at high scores, the rank ordering is well preserved, as evidenced by the near-identical ROC curves shown in Figure~\ref{fig:teacher_vs_student_roc}.
The largest absolute difference is 0.1 percentage points, observed for $A\to4\ell$. This agreement demonstrates that knowledge distillation effectively transfers the NF's anomaly-detection capability to a model with $325\times$ fewer parameters.
The performance is further quantified in Table~\ref{tab:student_full_precision_performance}, which reports the AUC together with the TPR at fixed FPR working points  from $10^{-2}$  to $10^{-5}$.

\begin{table*}[t]
  \centering
  \caption{Performance of the full-precision student model across the four BSM
  benchmarks. Values correspond to the nominal result; uncertainties are the
  RMS deviations obtained from four additional independent trainings relative to the nominal result}
  \label{tab:student_full_precision_performance}
  \resizebox{\textwidth}{!}{%
  \begin{tabular}{lccccc}
    \toprule
    & AUC (\%)
    & TPR @ $10^{-2}$ (\%)
    & TPR @ $10^{-3}$ (\%)
    & TPR @ $10^{-4}$ (\%)
    & TPR @ $10^{-5}$ (\%) \\
    \midrule
    $LQ\to b\tau$
    & $86.09 \pm 0.61$
    & $13.53 \pm 0.31$
    & $1.57 \pm 0.08$
    & $0.19 \pm 0.02$
    & $0.04 \pm 0.00$ \\

    $A\to 4\ell$
    & $90.66 \pm 0.18$
    & $37.01 \pm 2.12$
    & $12.45 \pm 1.34$
    & $7.52 \pm 0.46$
    & $3.10 \pm 0.23$ \\

    $h^{0}\to\tau\tau$
    & $74.00 \pm 0.93$
    & $11.26 \pm 0.33$
    & $2.70 \pm 0.14$
    & $0.41 \pm 0.02$
    & $0.05 \pm 0.01$ \\

    $h^{\pm}\to\tau\nu$
    & $89.33 \pm 0.79$
    & $20.98 \pm 1.54$
    & $2.70 \pm 0.16$
    & $0.29 \pm 0.01$
    & $0.05 \pm 0.00$ \\
    \bottomrule
  \end{tabular}%
  }
\end{table*}


\section{Model compression}
\label{sec:quantisation}

While the student model is compact in terms of architecture, it still operates in 32-bit floating-point arithmetic. FPGA deployment requires conversion to fixed-point arithmetic, which reduces both resource consumption (digital signal processing (DSP) slices, lookup tables (LUTs), and flip-flops (FFs)) and inference latency. Na\"ive post-training quantisation (PTQ), simply casting trained weights to lower precision after training, typically incurs significant accuracy loss at aggressive bit widths, as demonstrated in Ref.~\cite{Aarrestad:2021hls4mlcnn}. Quantisation-aware training (QAT) addresses this by simulating fixed-point arithmetic during the forward pass while retaining full-precision gradients via the straight-through estimator~\cite{Courbariaux:2015binaryconnect}, allowing the model to adapt to quantisation noise during optimisation.

We employ the \pquant framework~\cite{Niemi_2026} for QAT. \pquant is a neural network compression framework that provides quantised layers (\texttt{PQDense}, \texttt{PQActivation}) supporting:
\begin{itemize}[leftmargin=2em]
  \item Fixed-point quantisation of weights, biases, and activations with independently configurable integer and fractional bit widths;
  \item Multiple pruning methods with various pruning metrics;
  \item High-granularity quantisation~\cite{Sun:2024soe}, a QAT technique in which the optimal quantization of individual weights is learned during training;
  \item Configurable overflow modes (saturation, wrapping, symmetric saturation) and rounding modes.
\end{itemize}
All compression behaviour is controlled through a \texttt{PQConfig} configuration object, enabling systematic and reproducible scans over quantisation parameters. 
\pquant can be interfaced with FPGA porting tools such as  \alkaid~\cite{Sun:2025orx}, \hlsml~\cite{Schulte:2025mai}, and \aiefml~\cite{Danopoulos:2025tem}, so that the compressed models can be ported to FPGA hardware.

\subsection{Uniform bit-width quantisation}

The quantised student retains the architecture of the full-precision model,
with each dense layer replaced by its \pquant counterpart. To determine the
numerical precision required for FPGA deployment, we scan uniform bit widths
of 1, 2, 3, 4, 8, and 16 bits for the hidden-layer activations and weights.
At each scan point, the activation and weight precisions are varied
simultaneously, while the network input and output quantizers are kept fixed
to a signed 16-bit format with four integer and eleven fractional bits. The
higher input precision limits information loss when the standardised features
enter the network, while the 16-bit output provides sufficient resolution for
the continuous anomaly score.

The integer and fractional bit allocations used for the hidden-layer
activations and weights at each scan point are reported in
Table~\ref{tab:bitscan_config}. The activations are unsigned, whereas the
weight formats include a sign bit. The 1-bit and 2-bit configurations use
fixed-point representations and should therefore not be interpreted as strict
binary and ternary quantisation, respectively. Saturating arithmetic is used
throughout, such that values outside the representable range are clamped to
the nearest representable value rather than allowed to wrap around. \\

For each scan point, the quantised network is initialised from the same
trained full-precision student by copying the kernel and bias parameters of
each dense layer into the corresponding quantised layer. Each model is then
trained for 100 QAT epochs using the same SM training sample and teacher
tail-score targets defined in Eq.~\eqref{eq:ECDF}. The remaining training
configuration is identical to that of the full-precision student: we use the
MAE loss, the Adam optimizer with an initial learning rate of $10^{-3}$ and
a weight decay of $10^{-4}$, and a batch size of 1024. No additional
pretraining or post-training fine-tuning stages are applied. Pruning is
disabled to isolate the effect of numerical precision from that of network
sparsification.

\begin{table*}[t]
\centering
\caption{Data and weight bit allocation used in the uniform bit-width scan.
  Hidden-layer activations are unsigned, while the weight formats include an
  additional sign bit. The signed input and output quantizers are fixed to
  16 bits, with four integer and eleven fractional bits, throughout the scan}
\label{tab:bitscan_config}
\begin{tabular}{ccccc}
\toprule
Total bits & Data integer & Data fractional & Weight integer & Weight fractional \\
\midrule
1  & 1 & 0  &  0  & 0  \\
2  & 2 & 0  & 0 & 1  \\
3  & 3 & 0  & 0 & 2  \\
4  & 3 & 1  & 0 & 3  \\
8  & 3 & 5  & 0 & 7  \\
16  & 3 & 13  & 0 & 15 \\
 \bottomrule
\end{tabular}
\end{table*}

\begin{table*}[t]
  \centering
  \caption{AUC (\%) as a function of total bit width. Data and weight bit widths are varied simultaneously; input/output quantisation is fixed at 16 bits. The 32-bit column is the full-precision (floating-point) baseline model}
  \label{tab:bitscan}
  \begin{tabular}{lrrrrrrr}
    \toprule
    Signal              & 1-bit & 2-bit & 3-bit & 4-bit & 8-bit & 16-bit & 32-bit \\
    \midrule
    $LQ \to b\tau$      & 50.00 & 71.18 & 84.71 & 86.00 & 86.12 & 86.10 & 86.09 \\
    $A \to 4\ell$       & 50.00 & 81.44 & 88.41 & 90.51 & 90.61 & 90.68 & 90.66 \\
    $h^0 \to \tau\tau$  & 50.00 & 61.75 & 71.69 & 73.75 & 73.95 & 73.99 & 74.00 \\
    $h^\pm \to \tau\nu$ & 50.00 & 75.34 & 88.16 & 88.90 & 89.35 & 89.30 & 89.33 \\
    \bottomrule
  \end{tabular}%
\end{table*}

\DIFaddend \begin{figure}[h!]
  \centering
   \DIFaddbeginFL \includegraphics[width=\columnwidth]{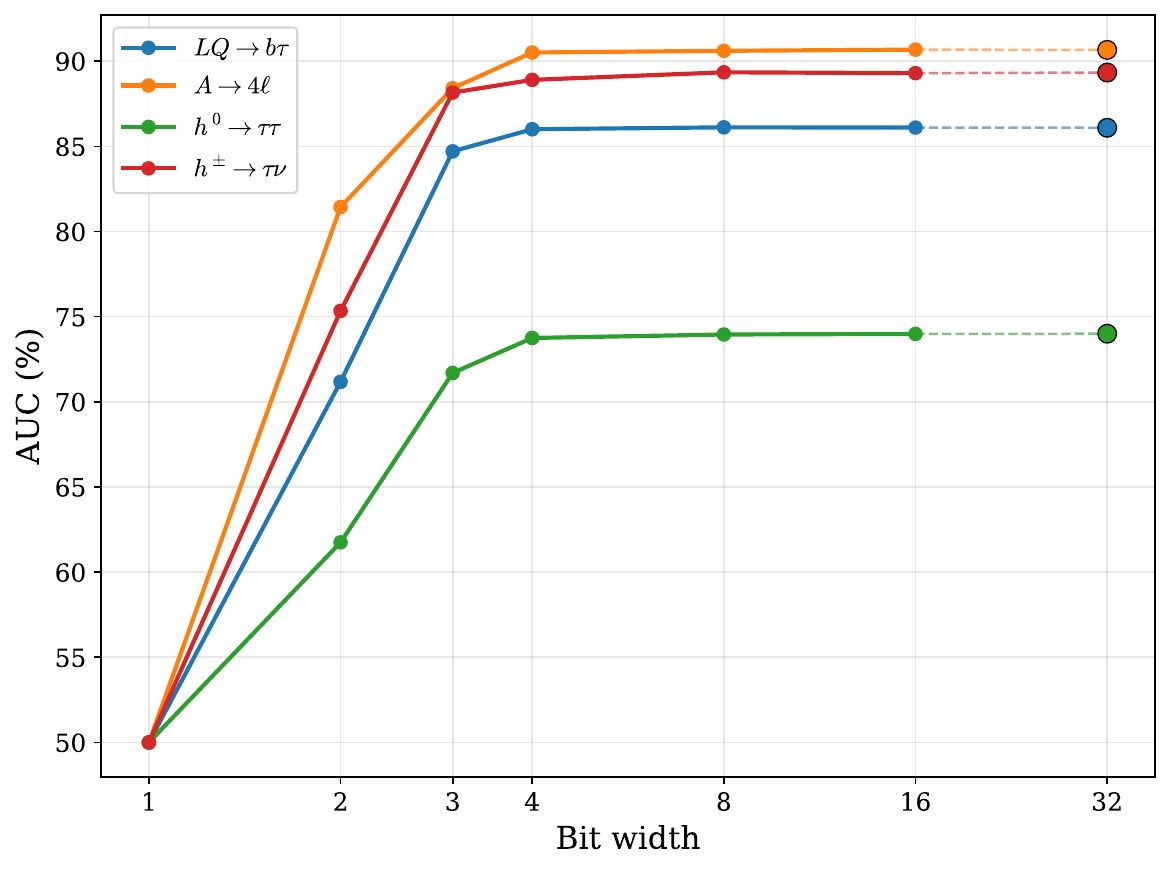}
  \DIFaddendFL \caption{AUC as a function of bit width for the four BSM benchmark signals.
     The 32-bit result corresponds to the full-precision (floating-point) baseline model}
  \label{fig:bitscan}
\end{figure}
The results of the uniform bit-width scan are reported in
Table~\ref{tab:bitscan} and illustrated in Figure~\ref{fig:bitscan}. The
1-bit configuration collapses to a constant score, yielding an AUC of 50\%
for all four benchmarks. The 2-bit model retains nontrivial discrimination
but loses 9--15 percentage points in AUC relative to the full-precision
student, depending on the benchmark. At 3 bits, the deficit is reduced to
1.2--2.3 percentage points, while at 4 bits it does not exceed
0.4 percentage points.
Based on this scan, we retain the uniform 8-bit QAT model as the reference
fixed-precision configuration. It reproduces the full-precision AUC to within
0.06 percentage points for every benchmark while substantially reducing the
arithmetic precision. Increasing the precision to 16 bits provides no
consistent performance improvement. In
Section~\ref{sec:hgq}, we investigate HGQ as an alternative approach that
learns the numerical precision at finer granularity. The uniform 8-bit and
HGQ configurations are subsequently compared in
Section~\ref{sec:quant_comparison}, using both the integrated AUC and the
performance at fixed low-FPR working points to select the nominal quantised
implementation.

\subsection{High-granularity quantisation}
\label{sec:hgq}
We also train the student using high-granularity quantisation (HGQ)~\cite{Coelho:2021hgq}, as implemented in \pquant. The HGQ network is warm-started by copying the kernels and biases of the trained full-precision student into the corresponding quantised layers. Its output remains linear, matching the tail-aware target $-\log_{10}(1-p)$. We initialise HGQ from the reference 8-bit format of Table~\ref{tab:bitscan}.
Training consists of 10 pretraining epochs, 100 HGQ epochs, and 20 fine-tuning epochs. The same full precision student architecture and training configuration are used, including the loss function, optimizer, learning rate, weight decay, and batch size. HGQ is applied with per-weight granularity, $\gamma=0$, and no explicit pruning.

\begin{figure}[t]
  \centering
  \includegraphics[width=\columnwidth]{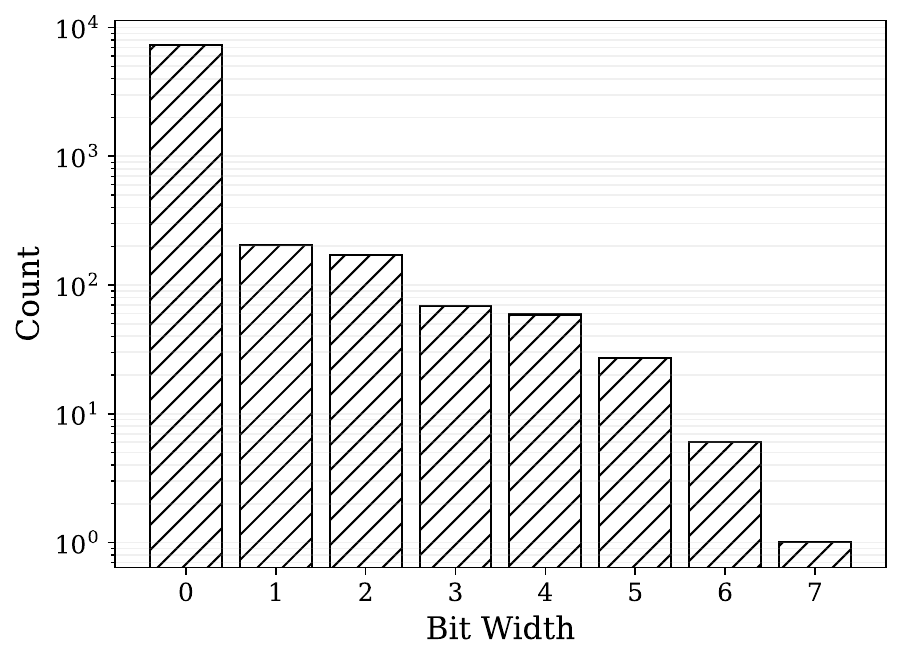}
  \caption{Distribution of learned weight bit-widths for the selected HGQ model at $\beta=10^{-6}$}
  \label{fig:bitwidth}
\end{figure}

We scan $\beta=5\times10^{-7},10^{-6},2\times10^{-6}$, and $5\times10^{-6}$, where $\beta$ controls the accuracy--complexity trade-off~\cite{chang2024mixedprecision}. A point is Pareto-optimal when no other configuration simultaneously has fewer effective bit operations (EBOPs) and an equal or larger mean AUC ratio. We found that only $\beta=10^{-6}$ and $5\times10^{-6}$ are Pareto-optimal. The $5\times10^{-7}$ point is dominated by $10^{-6}$, while $2\times10^{-6}$ is dominated by $5\times10^{-6}$.
We select $\beta=10^{-6}$ as the HGQ reference configuration because it
provides the largest mean AUC ratio, 0.9882, among the scanned points, while
reducing the arithmetic complexity to 17\,107 EBOPs.
Figure~\ref{fig:bitwidth} shows the learned bit-width distribution of the
kernel weights for the selected configuration. Of the 7\,808 kernel weights,
93.14\% are assigned a learned precision of zero bits (i.e., the weights are pruned), while the remaining
nonzero weights use precisions ranging from one to seven bits.

\subsection{Performance comparison and nominal configuration}
\label{sec:quant_comparison}

\begin{figure}[t]
  \centering
  \includegraphics[width=\columnwidth]{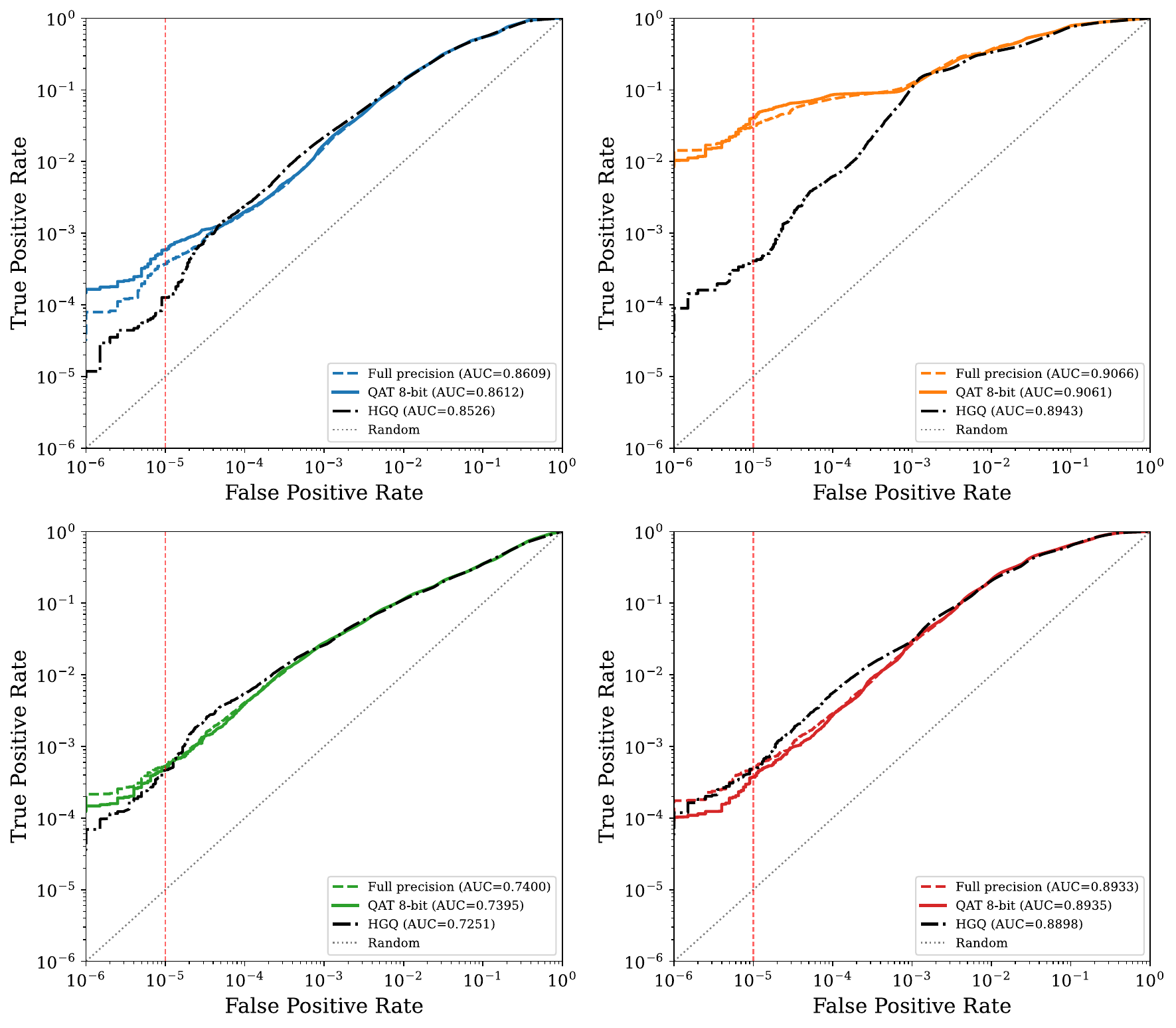}
\caption{ROC curves comparing the nominal full-precision student (dashed), the uniform 8-bit QAT model (solid), and the selected
HGQ model at $\beta=10^{-6}$ (black dash-dotted)}
  \label{fig:hgq_roc}
\end{figure}

Figure~\ref{fig:hgq_roc} compares the ROC curves of the full-precision
student, the reference uniform 8-bit QAT model, and the selected HGQ
configuration. Their AUC values and TPRs at fixed FPR working points are
reported in Table~\ref{tab:hgq_comparison}. The AUC provides an integrated
measure of discrimination across the full ROC curve, whereas the fixed-FPR
values directly probe the low-background tail relevant for the target
deployment.

\begin{table*}[t]
  \centering
  \caption{Performance comparison between the nominal full-precision student, the reference uniform 8-bit QAT model, and the selected HGQ model ($\beta=10^{-6}$). TPR values are obtained by direct interpolation of the original ROC points}
  \label{tab:hgq_comparison}
  \resizebox{\textwidth}{!}{%
  \begin{tabular}{llrrrrr}
    \toprule
    Signal & Model & AUC (\%) & TPR @ $10^{-2}$ (\%) & TPR @ $10^{-3}$ (\%) & TPR @ $10^{-4}$ (\%) & TPR @ $10^{-5}$ (\%) \\
    \midrule
    \multirow{3}{*}{$LQ\to b\tau$}
      & Student float32   & 86.09 & 13.53 & 1.57 & 0.19 & 0.04 \\
      & Uniform 8-bit QAT & 86.12 & 13.61 & 1.72 & 0.20 & 0.06 \\
      & HGQ               & 85.26 & 13.59 & 2.18 & 0.24 & 0.01 \\
    \midrule
    \multirow{3}{*}{$A\to4\ell$}
      & Student float32   & 90.66 & 37.01 & 12.45 & 7.52 & 3.10 \\
      & Uniform 8-bit QAT & 90.61 & 35.86 & 11.64 & 8.56 & 4.03 \\
      & HGQ               & 89.43 & 33.46 & 10.78 & 0.62 & 0.04 \\
    \midrule
    \multirow{3}{*}{$h^0\to\tau\tau$}
      & Student float32   & 74.00 & 11.26 & 2.70 & 0.41 & 0.05 \\
      & Uniform 8-bit QAT & 73.95 & 11.37 & 2.80 & 0.39 & 0.05 \\
      & HGQ               & 72.51 & 11.17 & 2.58 & 0.55 & 0.05 \\
    \midrule
    \multirow{3}{*}{$h^\pm\to\tau\nu$}
      & Student float32   & 89.33 & 20.98 & 2.70 & 0.29 & 0.05 \\
      & Uniform 8-bit QAT & 89.35 & 21.38 & 2.92 & 0.27 & 0.04 \\
      & HGQ               & 88.98 & 20.62 & 2.98 & 0.56 & 0.05 \\
    \bottomrule
  \end{tabular}%
  }
\end{table*}
The uniform 8-bit QAT model closely reproduces the full-precision student
across all four benchmarks, both in integrated AUC and at the fixed-FPR
working points. Its AUC differs from the floating-point reference by no more
than 0.06 percentage points, and no systematic degradation is observed in
the low-background tail.

The selected HGQ model provides a substantially lower-complexity
implementation and retains comparable integrated discrimination for three of
the four benchmarks. Its largest degradation is observed for
$A\to4\ell$: although its AUC is only 1.2 percentage points below that of
the full-precision student, its TPR decreases from 7.52\% to 0.62\% at an
FPR of $10^{-4}$ and from 3.10\% to 0.04\% at an FPR of $10^{-5}$. This
illustrates that similar integrated AUC values do not necessarily imply
equivalent performance in the sparsely populated low-background tail. At the
lowest working point, where an FPR of $10^{-5}$ corresponds to only 20
accepted background events in the evaluation sample, small differences
should nevertheless be interpreted with caution.

The uniform 8-bit QAT configuration is therefore selected as the nominal
quantised result of this study. It provides the most consistent performance
across all four benchmarks, including at the low-FPR working points, while
satisfying the FPGA resource and latency requirements. HGQ is retained as a
substantially lower-complexity alternative when minimising the hardware cost
is prioritised over uniform performance across the complete set of
benchmarks.

\section{FPGA implementation with \alkaid}
\label{sec:fpga}
\begin{table*}[t]
  \centering
  \caption{Out-of-context post-route timing and resource usage for the student models at different bit widths: $F_{\max}$ (maximum clock frequency), DSP, LUT, FF, and BRAM. The 16-bit design failed to route due to congestion, and the reported resource usage is obtained from post-place estimates}
  \label{tab:fpga_results}
  \begin{tabular}{crrrrrr}
    \toprule
    Bit width & $\mathrm{F}_{\max}$ [MHz] & Latency [ns] & DSP & LUT     & FF     & BRAM-18k \\
    \midrule
    3         & 221.1                     & 27.138       & 0   & 15,720  & 8,160  & 0        \\
    4         & 222.2                     & 27.000       & 0   & 20,954  & 6,867  & 0        \\
    8         & 211.8                     & 51.942       & 0   & 54,740  & 23,693 & 0        \\
    16$^{a}$  & N/A       & N/A          & 0   & 153,424 & 75,098 & 0        \\
    \bottomrule
  \end{tabular}
  \par\smallskip
  \raggedright\textsuperscript{a} Post-place estimate; the design failed to route
\end{table*}
For the Keras--RTL bit-accuracy validation, all standardised input
features are explicitly quantised to the signed Q4.11 fixed-point
format before being passed to either backend. This format comprises
one sign bit, four integer bits, and eleven fractional bits, for a
total width of 16 bits and a representable range of
$[-16,\,15.9995]$. Round-to-nearest with ties-to-even and saturation
on overflow are applied during this conversion. This prevents the two
backends from independently converting the original floating-point
inputs according to potentially different rounding and overflow
conventions. In particular, it avoids spurious Keras--RTL
discrepancies caused by saturation in the Keras input quantiser and
wrap-around during an implicit RTL cast, ensuring that both backends
are evaluated using identical Q4.11 input representations. 
The quantised student models are lowered with \alkaid, a successor compiler of \dafml~\cite{Sun:2025orx} for low-latency static-dataflow kernels that symbolically traces the fixed-point model and emits pipelined Verilog RTL directly. During tracing, affine-arithmetic precision propagation, with interval arithmetic as a fallback, tracks intermediate bit growth and narrows each value only to the width required by its inferred fixed-point range. No requantisation or truncation not explicitly defined by the quantised model is introduced in the process, making the generated RTL bit-exact with respect to the quantised fixed-point computation. Consistent with \dafml, constant-matrix-vector multiplications are lowered to multiplier-free distributed arithmetic. Sparsity-aware LUT6 packing that groups one-bit operands, and target-aware ternary-adder fusion are enabled to further reduce the resource footprint while preserving the bit-exact arithmetic semantics.

\subsection{Synthesis configuration}

The 2-, 3-, 4-, 8-, and 16-bit student models are compiled with the Verilog backend of \alkaid targeting Xilinx Virtex UltraScale+ VU13P (\texttt{xcvu13p-flga2577-2-e}) at a target frequency of 200 MHz. Table~\ref{tab:fpga_results} shows the on-chip resource usage and  timing  measurements for out-of-context  designs. The 16-bit design failed to route  because of  excessive congestion,  so its resource usage is reported from post-place estimates  rather than post-route  results. The 1-bit model is omitted because it collapses to a constant score, while the strongly degraded  2-bit model is not considered a viable deployment candidate. All routed designs meet the specified timing target of 200 MHz without using DSP slices or block random-access memory (BRAM) blocks.

\section{Comparison with existing approaches}
\label{sec:discussion}

\begin{table*}[t]
  \centering
  \caption{Comparison of the AUC and TPR at a fixed FPR of $10^{-5}$
  across different anomaly-detection models evaluated on the same four BSM
  benchmarks. Floating-point and quantised implementations are shown
  separately. The PTQ and HGQ configurations of
  Ref.~\cite{Vaselli:2025fad} employ different quantisation schemes from the
  uniform 8-bit QAT configuration used in this work. All values are given
  in percent}
  \label{tab:comparison_performance}
  \resizebox{\textwidth}{!}{%
  \begin{tabular}{l cccc cccc l}
    \toprule
    \multirow{2}{*}{Model}
    & \multicolumn{4}{c}{AUC [\%]}
    & \multicolumn{4}{c}{TPR @ FPR $10^{-5}$ [\%]}
    & \multirow{2}{*}{Ref.} \\
    \cmidrule(lr){2-5}
    \cmidrule(lr){6-9}
    & $LQ\to b\tau$
    & $A\to4\ell$
    & $h^\pm\to\tau\nu$
    & $h^0\to\tau\tau$
    & $LQ\to b\tau$
    & $A\to4\ell$
    & $h^\pm\to\tau\nu$
    & $h^0\to\tau\tau$
    & \\
    \midrule

    \multicolumn{10}{l}{\textbf{Floating-point models}} \\
    \addlinespace[2pt]

    Flow $v_t$
    & 80 & 82 & 84 & 68
    & 0.04 & 2.8 & 0.04 & 0.06
    & \cite{Vaselli:2025fad} \\

    Flow ODE
    & 80 & 88 & 86 & 69
    & 0.04 & 3.8 & 0.04 & 0.05
    & \cite{Vaselli:2025fad} \\

    VAE (retrained)
    & 59 & 72 & 63 & 57
    & 0.02 & 2.4 & 0.02 & 0.04
    & \cite{Vaselli:2025fad} \\

    RealNVP Teacher (ours)
    & 86 & 91 & 89 & 74
    & 0.05 & 4.1 & 0.05 & 0.07
    & This work \\

    Student (float32, ours) 
    & 86 & 91 & 89 & 74
    & 0.04 & 3.1 & 0.05 & 0.05
    & This work \\

    \midrule

    \multicolumn{10}{l}{\textbf{Quantised models}} \\
    \addlinespace[2pt]

    Flow $v_t$ (PTQ)
    & 75 & 81 & 81 & 65
    & 0.03 & 3.6 & 0.04 & 0.06
    & \cite{Vaselli:2025fad} \\

    Flow $v_t$ (HGQ)
    & 77 & 86 & 82 & 66
    & 0.04 & 3.4 & 0.05 & 0.06
    & \cite{Vaselli:2025fad} \\

    Student (8-bit QAT, ours)
    & 86 & 91 & 89 & 74
    & 0.06 & 4.0 & 0.04 & 0.05
    & This work \\

    \bottomrule
  \end{tabular}%
  }
\end{table*}

Table~\ref{tab:comparison_performance} places our results in the context of
existing approaches evaluated on the same benchmark. The retrained VAE of
Ref.~\cite{Vaselli:2025fad} achieves AUC values of only 57--72\% across the
four signals. The two
floating-point CNF-based approaches of Ref.~\cite{Vaselli:2025fad} perform
substantially better. Flow~$v_t$ avoids ODE integration by using the squared
norm of the vector field as its anomaly score, while Flow~ODE integrates the
full ODE, yielding AUC values of 68--84\% and 69--88\%, respectively.
The pipeline presented in this work achieves higher AUC values than all the
reference models across the four benchmarks. The RealNVP teacher, comprising
approximately 2.6M parameters, and the distilled float32 student achieve
statistically compatible discrimination within the observed variation across
independent trainings. The student contains only 7{,}937 parameters, while
retaining most of the discrimination power of the substantially larger
teacher. Although this corresponds to approximately four times the number of
parameters used by the CNF models of Ref.~\cite{Vaselli:2025fad}, the student
surpasses both Flow~$v_t$ and Flow~ODE on every benchmark.
This improvement cannot be attributed solely to model size. To investigate this point, we also trained larger flow-matching architectures following the approach of Ref.~\cite{Vaselli:2025fad}. Consistent with the observations reported in that work, increasing the network capacity alone did not lead to comparable gains in anomaly-detection performance.
  This suggests that the standard flow-matching objective may constrain anomaly-detection performance through its connection to the recovery of the $L_2$ norm, thereby limiting the benefits of additional representational power unless trained on improved objective functions.
  In contrast, the distilled student benefits from the richer supervision provided by the $\sim$2.6M-parameter discrete-flow teacher,
  whose training objective is not explicitly constrained to reproduce the norm operation. Consequently, the student inherits much of the teacher's discrimination power while remaining compact enough for FPGA
  deployment.

For a direct comparison between hardware-oriented implementations,
Table~\ref{tab:comparison_performance} also includes the PTQ and HGQ versions
of Flow~$v_t$ reported in Ref.~\cite{Vaselli:2025fad}. These configurations
are not directly equivalent to our uniform 8-bit QAT model: the PTQ
implementation uses different fixed-point precisions for weights and
operations, while HGQ assigns heterogeneous per-parameter precisions. Nevertheless, the 8-bit distilled student achieves higher
AUC values than both quantised Flow~$v_t$ implementations across all four
benchmarks. Moreover, 8-bit QAT changes the student AUC by no more than
0.06 percentage points relative to its full-precision counterpart, confirming
that the learned mapping is well suited to fixed-point arithmetic.

To complement the integrated AUC comparison,
Table~\ref{tab:comparison_performance} also reports the TPR at the stringent
FPR working point of $10^{-5}$. The distilled students remain competitive
with the reference approaches across all four benchmarks. In particular, the
8-bit student reaches a nominal TPR of 4.0\% for $A\to4\ell$, exceeding the
values reported for the floating-point Flow~$v_t$, Flow~ODE, Flow~$v_t$
(PTQ), and Flow~$v_t$ (HGQ) models. The close agreement between the float32
and 8-bit students indicates that quantisation preserves the discrimination
performance not only globally, as measured by the AUC, but also in the
sparsely populated low-background tail.

\FloatBarrier

\section{Summary}
\label{sec:conclusions}

We present a pipeline for unsupervised anomaly detection at the CMS Level-1 Trigger based on knowledge distillation of a normalising flow. The teacher model is a RealNVP architecture, whose anomaly score is given by the exact negative log-likelihood. This provides a more principled quantity than the reconstruction errors or ELBO-based approximations used in previously deployed autoencoder approaches.
A compact student MLP with 7{,}937 parameters is trained to regress the teacher’s anomaly scores. The student reproduces the teacher’s discrimination power across four BSM benchmarks to within 0.1 percentage points in AUC at a compression factor of $325\times$. The student also outperforms the previously published continuous normalising flow~\cite{Vaselli:2025fad} deployed directly at the L1T,  while remaining compact enough for FPGA implementation.
This result indicates that the performance gain stems from the richer supervision provided by distilling a high-capacity, exact-likelihood teacher, as opposed to training a compact model directly on a hardware-friendly proxy score such as the squared vector-field norm of Ref.~\cite{Vaselli:2025fad}.
Finally, quantisation-aware training with \pquant reduces the student model to 8-bit precision with AUC changes below 0.06 percentage points. The 3-bit model remains functional but loses 1.2--2.3 percentage points relative to the full precision student, whereas the 4-bit deficit is at most 0.4 percentage points.
The 3-, 4-, and 8-bit models are implemented as direct Verilog RTL using \alkaid. The routed implementations use no DSPs or block RAM and achieve latencies of around 27--52 ns. 

Based on our results, anomaly detection based on normalising flows emerges as a strong candidate to replace autoencoders in the ATLAS and CMS L1Ts in the High-Luminosity phase of the LHC.


\section*{Acknowledgements}
A first draft of this paper was generated with the assistance of Anthropic’s Claude, based on existing code, a technical document, and a set of publications on related FPGA toolflows. The draft was subsequently reviewed and edited by the authors. OpenAI ChatGPT has been used for a final editorial check.
Anthropic's Claude and OpenAI ChatGPT were also used, under the authors’ supervision, to assist with code development. All AI-assisted code was reviewed and validated by the authors.
While we were finalizing our paper, we became aware of Ref.~\cite{Tahseen:2026uaa}, which explores similar ideas with similar findings.

\begingroup
\section*{Statements and Declarations}

\subsection*{Funding}

Roope Oskari Niemi acknowledges funding from the Eric \& Wendy Schmidt Fund
for Strategic Innovation through the CERN Next Generation Triggers project
under Grant Agreement No.~SIF-2023-004.\\

\noindent{Chang Sun is partially supported by United States DoE (grant numbers
DE-SC0011925, DE-FOA-0002705) and NSF (grant numbers PHY240298, PHY2117997).}

\subsection*{Competing interests}

The authors have no relevant financial or non-financial interests to disclose.

\subsection*{Data availability}

The datasets analysed in this study are publicly available on Zenodo
and are cited in Refs.~\cite{zenodo:lq,zenodo:a4l,zenodo:h0tt,zenodo:hctaunu,Aarrestad:2021v2}.

\subsection*{Code availability}

The custom code developed for this study is not publicly available but is available from the corresponding author upon reasonable request.
\endgroup


\bibliography{bibliography_inspire_style_1}

@article{ATLAS:2008xda,
  author        = "{ATLAS Collaboration}",
  title         = "{The ATLAS Experiment at the CERN Large Hadron Collider}",
  doi           = "10.1088/1748-0221/3/08/S08003",
  url           = "https://doi.org/10.1088/1748-0221/3/08/S08003",
  journal       = "JINST",
  volume        = "3",
  pages         = "S08003",
  year          = "2008"

}

@misc{Aarrestad:2021v2,
  author    = "Aarrestad, Thea and Govorkova, Ekaterina and Ngadiuba, Jennifer and Puljak, Ema and Pierini, Maurizio and Wozniak, Kinga Anna",
  title     = "Unsupervised New Physics Detection at 40 MHz: Training Dataset",
  version   = "v2",
  publisher = "Zenodo",
  doi       = "10.5281/zenodo.5046428",
  year      = "2021"
}

@article{CMS:2008xjf,
  author        = "{CMS Collaboration}",
  title         = "{The CMS Experiment at the CERN LHC}",
  doi           = "10.1088/1748-0221/3/08/S08004",
  url           = "https://doi.org/10.1088/1748-0221/3/08/S08004",
  journal       = "JINST",
  volume        = "3",
  pages         = "S08004",
  year          = "2008"
}

@article{Tahseen:2026uaa,
    author = "Tahseen, Tara P. A. and Clarke Hall, Noah and Konstantinidis, Nikolaos and Su{\'a}rez, Paula Mart{\'\i}nez",
    title = "{Distilling Normalizing Flows for Real-Time Anomaly Detection at the LHC}",
    eprint = "2608.19912",
    archivePrefix = "arXiv",
    primaryClass = "hep-ex",
    month = "8",
    year = "2026"
}

@article{CMS:2020cmk,
  author        = "{CMS Collaboration}",
  title         = "{Performance of the CMS Level-1 trigger in proton-proton collisions at $\sqrt{s}=13$ TeV}",
  doi           = "10.1088/1748-0221/15/10/P10017",
  url           = "https://doi.org/10.1088/1748-0221/15/10/P10017",
  journal       = "JINST",
  volume        = "15",
  number        = "10",
  pages         = "P10017",
  year          = "2020"
}

@techreport{CMS:L1TDR,
  author        = "{CMS Collaboration}",
  title         = "{The Phase-2 Upgrade of the CMS Level-1 Trigger}",
  reportNumber  = "CERN-LHCC-2020-004, CMS-TDR-021",
  institution   = "{CERN}",
  url           = "https://cds.cern.ch/record/2714892",
  year          = "2020",
  note          = {\href{https://cds.cern.ch/record/2714892}{\texttt{cds:2714892}}}
}

@article{Kasieczka:2021xcg,
  author        = "Kasieczka, Gregor and others",
  title         = "{The LHC Olympics 2020: a community challenge for anomaly detection in high energy physics}",
  eprint        = "2101.08320",
  archivePrefix = "arXiv",
  primaryClass  = "hep-ph",
  doi           = "10.1088/1361-6633/ac36b9",
  url           = "https://doi.org/10.1088/1361-6633/ac36b9",
  journal       = "{Rep. Prog. Phys.}",
  volume        = "84",
  number        = "12",
  pages         = "124201",
  year          = "2021"
}

@article{Aarrestad:2021oeb,
  author        = "Aarrestad, T. and others",
  title         = "{The Dark Machines Anomaly Score Challenge}",
  eprint        = "2105.14027",
  archivePrefix = "arXiv",
  primaryClass  = "hep-ph",
  doi           = "10.21468/SciPostPhys.12.1.043",
  url           = "https://doi.org/10.21468/SciPostPhys.12.1.043",
  journal       = "SciPost Phys.",
  volume        = "12",
  number        = "1",
  pages         = "043",
  year          = "2022"
}

@article{Kingma:2013vae,
  author        = "Kingma, Diederik P. and Welling, Max",
  title         = "{Auto-Encoding Variational Bayes}",
  eprint        = "1312.6114",
  archivePrefix = "arXiv",
  primaryClass  = "stat.ML",
  url           = "https://arxiv.org/abs/1312.6114",
  year          = "2013"
}

@article{Cerri:2018anp,
  author        = "Cerri, Olmo and Nguyen, Thong Q. and Pierini, Maurizio and Spiropulu, Maria and Vlimant, Jean-Roch",
  title         = "{Variational autoencoders for new physics mining at the Large Hadron Collider}",
  eprint        = "1811.10276",
  archivePrefix = "arXiv",
  primaryClass  = "hep-ex",
  doi           = "10.1007/JHEP05(2019)036",
  url           = "https://doi.org/10.1007/JHEP05(2019)036",
  journal       = "JHEP",
  volume        = "05",
  pages         = "036",
  year          = "2019"
}

@article{Knapp:2020adversarial,
  author        = "Knapp, Oliver and others",
  title         = "{Adversarially Learned Anomaly Detection on CMS Open Data: re-discovering the top quark}",
  eprint        = "2005.01598",
  archivePrefix = "arXiv",
  primaryClass  = "hep-ex",
  url           = "https://arxiv.org/abs/2005.01598",
  year          = "2020"
}

@article{Govorkova:2021utb,
  author        = "Govorkova, Ekaterina and others",
  title         = "{Autoencoders on FPGAs for real-time, unsupervised new physics detection at 40 MHz at the LHC}",
  eprint        = "2108.03986",
  archivePrefix = "arXiv",
  primaryClass  = "physics.ins-det",
  doi           = "10.1038/s42256-022-00441-3",
  url           = "https://doi.org/10.1038/s42256-022-00441-3",
  journal       = "{Nat. Mach. Intell.}",
  volume        = "4",
  number        = "2",
  pages         = "154--161",
  year          = "2022"
}

@article{Gandrakota:2024axoltl,
  author        = "Gandrakota, Adarsh",
  title         = "{Real-time Anomaly Detection at the L1 Trigger of CMS Experiment}",
  eprint        = "2411.19506",
  archivePrefix = "arXiv",
  primaryClass  = "hep-ex",
  url           = "https://arxiv.org/abs/2411.19506",
  year          = "2024"
}

@article{Hinton:2015distill,
  author        = "Hinton, Geoffrey and Vinyals, Oriol and Dean, Jeff",
  title         = "{Distilling the Knowledge in a Neural Network}",
  eprint        = "1503.02531",
  archivePrefix = "arXiv",
  primaryClass  = "stat.ML",
  url           = "https://arxiv.org/abs/1503.02531",
  year          = "2015"
}

@misc{Cohen:2025gelato,
  author        = "Cohen, M. M. and others",
  title         = "{GELATO: a Generic Event-Level Anomalous Trigger Option for ATLAS}",
  howpublished  = "CERN CDS record 2938881",
  url           = "https://cds.cern.ch/record/2938881",
  year          = "2025"
}

@article{Vaselli:2025fad,
    author = "Vaselli, Francesco and Sun, Chang and Aarrestad, Thea and Danopoulos, Dimitrios and Niemi, Roope Oskari and Glowacki, Maciej Mikolaj and Govorkova, Katya and Loncar, Vladimir and Pantaleo, Felice and Pierini, Maurizio",
    title = "{It{\textquoteright}s not a FAD: first demonstration of flows for unsupervised anomaly detection at 40{\,}MHz for use at the Large Hadron Collider}",
    eprint = "2508.11594",
    archivePrefix = "arXiv",
    primaryClass = "hep-ex",
    doi = "10.1088/2632-2153/ae51dd",
    url = "https://doi.org/10.1088/2632-2153/ae51dd",
    journal = "{Mach. Learn.: Sci. Technol.}",
    volume = "7",
    number = "2",
    pages = "025052",
    year = "2026"
}

@article{Pol:2023ots,
  author        = "Pol, Adrian Alan and others",
  title         = "{Knowledge Distillation for Anomaly Detection}",
  eprint        = "2310.06047",
  archivePrefix = "arXiv",
  primaryClass  = "cs.LG",
  url           = "https://arxiv.org/abs/2310.06047",
  year          = "2023"
}

@article{Dinh:2016realnvp,
  author        = "Dinh, Laurent and Sohl-Dickstein, Jascha and Bengio, Samy",
  title         = "{Density estimation using Real NVP}",
  eprint        = "1605.08803",
  archivePrefix = "arXiv",
  primaryClass  = "cs.LG",
  url           = "https://arxiv.org/abs/1605.08803",
  year          = "2016"
}

@article{Papamakarios:2019fms,
    author = "Papamakarios, George and Nalisnick, Eric and Rezende, Danilo Jimenez and Mohamed, Shakir and Lakshminarayanan, Balaji",
    title = "{Normalizing Flows for Probabilistic Modeling and Inference}",
    eprint = "1912.02762",
    archivePrefix = "arXiv",
    primaryClass = "stat.ML",
    doi = "10.5555/3546258.3546315",
    url = "https://doi.org/10.5555/3546258.3546315",
    journal = "{J. Mach. Learn. Res.}",
    volume = "22",
    number = "1",
    pages = "2617--2680",
    year = "2021"
}

@misc{chang2024mixedprecision,
  author       = {C. Sun and others},
  title        = {Gradient-based Automatic Per-Weight Mixed Precision Quantization for Neural Networks On-Chip},
  year         = {2024},
  doi          = {10.7907/HQ8JD-RHG30}
}

@misc{PQuantML,
  author        = "Niemi, R. and others",
  title         = "{PQuantML: A framework for compressing neural networks via pruning and quantization}",
  howpublished  = "CERN Next Generation Triggers",
  url           = "https://next-generation-triggers.web.cern.ch",
  year          = "2025"
}

@article{Niemi_2026,
doi = {10.1088/2632-2153/ae94e3},
url = {https://doi.org/10.1088/2632-2153/ae94e3},
year = {2026},
month = {sep},
publisher = {IOP Publishing},
volume = {7},
number = {5},
pages = {055007},
author = {Niemi, Roope and Petrovych, Anastasiia and Ranjan Das, Arghya and Lupi, Enrico and Sun, Chang and Danopoulos, Dimitrios and Helbing, Marlon Joshua and Liu, Mia and Dittmeier, Sebastian and Kagan, Michael and Loncar, Vladimir and Pierini, Maurizio},
title = {{{PQuantML}: a tool for end-to-end hardware-aware model compression}},
journal = {{Mach. Learn.: Sci. Technol.}}
}

@article{Aarrestad:2021hls4mlcnn,
  author        = "Aarrestad, Thea and others",
  title         = "{Fast convolutional neural networks on FPGAs with hls4ml}",
  eprint        = "2101.05108",
  archivePrefix = "arXiv",
  primaryClass  = "physics.ins-det",
  doi           = "10.1088/2632-2153/ac0ea1",
  url           = "https://doi.org/10.1088/2632-2153/ac0ea1",
  journal       = "{Mach. Learn.: Sci. Technol.}",
  volume        = "2",
  number        = "4",
  pages         = "045015",
  year          = "2021"
}

@article{Kingma:2014adam,
  author        = "Kingma, Diederik P. and Ba, Jimmy",
  title         = "{Adam: A Method for Stochastic Optimization}",
  eprint        = "1412.6980",
  archivePrefix = "arXiv",
  primaryClass  = "cs.LG",
  url           = "https://arxiv.org/abs/1412.6980",
  year          = "2014"
}

@article{Coelho:2021hgq,
  author        = "Coelho, Claudionor N. and others",
  title         = "{Automatic heterogeneous quantization of deep neural networks for low-latency inference on the edge for particle detectors}",
  eprint        = "2006.10159",
  archivePrefix = "arXiv",
  primaryClass  = "physics.ins-det",
  doi           = "10.1038/s42256-021-00356-5",
  url           = "https://doi.org/10.1038/s42256-021-00356-5",
  journal       = "{Nat. Mach. Intell.}",
  volume        = "3",
  pages         = "675--686",
  year          = "2021"
}

@article{Schulte:2025mai,
  author        = "Schulte, Jan-Frederik and others",
  title         = "{hls4ml: A Flexible, Open-Source Platform for Deep Learning Acceleration on Reconfigurable Hardware}",
  eprint        = "2512.01463",
  archivePrefix = "arXiv",
  primaryClass  = "cs.AR",
  reportNumber  = "FERMILAB-PUB-25-0890-CSAID-ETD-PPD",
  url           = "https://arxiv.org/abs/2512.01463",
  month         = "12",
  year          = "2025"
}

@article{Sun:2025orx,
  author        = "Sun, Chang and Que, Zhiqiang and Loncar, Vladimir and Luk, Wayne and Spiropulu, Maria",
  title         = "{da4ml: Distributed Arithmetic for Real-time Neural Networks on FPGAs}",
  eprint        = "2507.04535",
  archivePrefix = "arXiv",
  primaryClass  = "cs.AR",
  doi           = "10.1145/3777387",
  url           = "https://doi.org/10.1145/3777387",
  month         = "7",
  year          = "2025"
}

@article{Danopoulos:2025tem,
  author        = "Danopoulos, Dimitrios and Lupi, Enrico and Sun, Chang and Dittmeier, Sebastian and Kagan, Michael and Loncar, Vladimir and Pierini, Maurizio",
  title         = "{AIE4ML: An End-to-End Framework for Compiling Neural Networks for the Next Generation of AMD AI Engines}",
  eprint        = "2512.15946",
  archivePrefix = "arXiv",
  primaryClass  = "cs.LG",
  url           = "https://arxiv.org/abs/2512.15946",
  month         = "12",
  year          = "2025"
}

@article{Courbariaux:2015binaryconnect,
  author        = "Courbariaux, Matthieu and Bengio, Yoshua and David, Jean-Pierre",
  title         = "{BinaryConnect: Training Deep Neural Networks with binary weights during propagations}",
  eprint        = "1511.00363",
  archivePrefix = "arXiv",
  primaryClass  = "cs.LG",
  url           = "https://arxiv.org/abs/1511.00363",
  journal       = "Adv. Neural Inf. Process. Syst.",
  volume        = "28",
  pages         = "3123--3131",
  year          = "2015"
}

@article{Govorkova:2021dataset,
  author        = "Govorkova, Ekaterina and others",
  title         = "{LHC physics dataset for unsupervised New Physics detection at 40 MHz}",
  eprint        = "2107.02157",
  archivePrefix = "arXiv",
  primaryClass  = "physics.data-an",
  url           = "https://arxiv.org/abs/2107.02157",
  year          = "2021"
}

@misc{zenodo:lq,
  author        = "Aarrestad, T. and others",
  title         = "Unsupervised New Physics detection at 40 MHz: $LQ \to b\tau$ Signal Benchmark Dataset",
  doi           = "10.5281/zenodo.7152599",
  publisher     = "Zenodo",
  year          = "2021"
}

@misc{zenodo:a4l,
  author        = "Aarrestad, T. and others",
  title         = "Unsupervised New Physics detection at 40 MHz: $A \to 4\ell$ Signal Benchmark Dataset",
  doi           = "10.5281/zenodo.5046445",
  publisher     = "Zenodo",
  year          = "2021"
}

@misc{zenodo:h0tt,
  author        = "Aarrestad, T. and others",
  title         = "Unsupervised New Physics detection at 40 MHz: $h^0 \to \tau\tau$ Signal Benchmark Dataset",
  doi           = "10.5281/zenodo.5061632",
  publisher     = "Zenodo",
  year          = "2021"
}

@misc{zenodo:hctaunu,
  author        = "Aarrestad, T. and others",
  title         = "Unsupervised New Physics detection at 40 MHz: $h^\pm \to \tau\nu$ Signal Benchmark Dataset",
  doi           = "10.5281/zenodo.5061687",
  publisher     = "Zenodo",
  year          = "2021"
}

@inproceedings{Sun:2024soe,
    author = "Sun, Chang and Que, Zhiqiang and {\r{A}}rrestad, Thea and Loncar, Vladimir and Ngadiuba, Jennifer and Luk, Wayne and Spiropulu, Maria",
    title = "{HGQ: High Granularity Quantization for Real-time Neural Networks on FPGAs}",
    booktitle = "The 2026 ACM/SIGDA International Symposium on Field Programmable Gate Arrays",
    eprint = "2405.00645",
    archivePrefix = "arXiv",
    primaryClass = "cs.LG",
    reportNumber = "FERMILAB-PUB-24-0213-CMS, CaltechAUTHORS:10.7907/hq8jd-rhg30",
    doi = "10.1145/3748173.3779200",
    publisher = "Association for Computing Machinery",
    address = "New York, NY, USA",
    pages = "79--91",
    month = "2",
    year = "2026"
}

@article{Heimel:2018mkt,
    author = "Heimel, Theo and Kasieczka, Gregor and Plehn, Tilman and Thompson, Jennifer M.",
    title = "{QCD or What?}",
    eprint = "1808.08979",
    archivePrefix = "arXiv",
    primaryClass = "hep-ph",
    doi = "10.21468/SciPostPhys.6.3.030",
    url = "https://doi.org/10.21468/SciPostPhys.6.3.030",
    journal = "SciPost Phys.",
    volume = "6",
    number = "3",
    pages = "030",
    year = "2019"
}

@article{Farina:2018fyg,
    author = "Farina, Marco and Nakai, Yuichiro and Shih, David",
    title = "{Searching for New Physics with Deep Autoencoders}",
    eprint = "1808.08992",
    archivePrefix = "arXiv",
    primaryClass = "hep-ph",
    doi = "10.1103/PhysRevD.101.075021",
    url = "https://doi.org/10.1103/PhysRevD.101.075021",
    journal = "Phys. Rev. D",
    volume = "101",
    number = "7",
    pages = "075021",
    year = "2020"
}

@article{Fraser:2021lxm,
    author = "Fraser, Katherine and Homiller, Samuel and Mishra, Rashmish K. and Ostdiek, Bryan and Schwartz, Matthew D.",
    title = "{Challenges for unsupervised anomaly detection in particle physics}",
    eprint = "2110.06948",
    archivePrefix = "arXiv",
    primaryClass = "hep-ph",
    doi = "10.1007/JHEP03(2022)066",
    url = "https://doi.org/10.1007/JHEP03(2022)066",
    journal = "JHEP",
    volume = "03",
    pages = "066",
    year = "2022"
}

\end{document}